\documentclass[11pt,a4paper]{article}
\usepackage{amsmath,amsthm,amssymb,epsfig,latexsym}
\usepackage{epsfig}
\usepackage[T1]{fontenc}    
\usepackage{graphics}
\usepackage{graphicx}
\usepackage{pstricks,pst-coil,pst-fill,pst-plot}
\newcommand{\be}[1]{\begin{equation}\label{#1}}
\newcommand{\ba}[1]{\begin{multline}\label{#1}}
\newcommand{\ee}{\end{equation}}
\newcommand{\ea}{\end{eqnarray}}

\newcommand{\num}{\\\rule{0pt}{20pt}}
\newcommand{\numa}[1]{\\\rule{0pt}{#1pt}}
\newcommand{\dis}{\displaystyle}

\newcommand{\Rp}{r^{\scriptscriptstyle +}}
\newcommand{\Rm}{r^{\scriptscriptstyle -}}
\newcommand{\Rpm}{r^{\scriptscriptstyle \pm}}
\newcommand{\TRpm}{{\tilde r}^{\scriptscriptstyle \pm}}
\newcommand{\hqp}{\hat s^{\scriptscriptstyle +}}
\newcommand{\hqm}{\hat s^{\scriptscriptstyle -}}
\newcommand{\hqpm}{\hat s^{\scriptscriptstyle \pm}}
\newcommand{\hqmp}{\hat s^{\scriptscriptstyle \mp}}

\newcommand{\moy}[1]{\ensuremath{\langle #1 \rangle}_T}

 \makeatletter
 \@addtoreset{equation}{section}
 \makeatother
 
\begin{document}

\begin{flushright}

LPENSL-TH-02/11

DESY 10-254
\end{flushright}
\par \vskip .1in \noindent

\vspace{24pt}

\begin{center}
\begin{LARGE}
\vspace*{1cm} {Correlation functions of one-dimensional bosons\\ at
low temperature}
\end{LARGE}

\vspace{30pt}

\begin{large}

{\bf K.~K.~Kozlowski}\footnote[1]{DESY, Hamburg, Deutschland,
 karol.kajetan.kozlowski@desy.de},~~
{\bf J.~M.~Maillet}\footnote[2]{ Laboratoire de Physique, UMR 5672
du CNRS, ENS Lyon,  France,
 maillet@ens-lyon.fr},~~
{\bf N.~A.~Slavnov}\footnote[3]{ Steklov Mathematical Institute,
Moscow, Russia, nslavnov@mi.ras.ru}

\end{large}

\vspace{30pt}

\vspace{1cm}
\parbox{12cm}{\small .}

\vspace{20pt}

\centerline{\bf Abstract} \vspace{1cm}
\parbox{12cm}{\small  We consider the low-temperature limit of the
long-distance asymptotic behavior of the finite temperature
density-density correlation function in the one-dimensional Bose gas
derived recently in the algebraic Bethe Ansatz framework. Our
results confirm the predictions based on the Luttinger liquid and
conformal field theory approaches. We also demonstrate that the
amplitudes arising in this asymptotic expansion at low-temperature
coincide with the amplitudes associated with the so-called critical
form factors. }
\end{center}

\vspace{20pt}

\section{Introduction\label{INTR}}

The model of one-dimensional bosons interacting through a two-body $\delta$-function potential
is described by the Quantum
Nonlinear Schr\"odinger equation (QNLS model). Recently
\cite{KozMS10}, the Lagrange series method was applied in the framework of the algebraic Bethe ansatz to the derivation of the long-distance asymptotic expansion of the finite
temperature density-density correlation function in this model. In the
present paper, we consider the low-temperature limit of this result.

The QNLS model is given by the Hamiltonian
 \begin{equation}\label{8-Ham}
 H=\int\limits_0^L\!\left(
 \partial_x\Psi^\dagger\partial_x\Psi
 +
 c\Psi^\dagger\Psi^\dagger\Psi\Psi-h\Psi^\dagger\Psi\right)\,d x\, .
 \end{equation}
Here $\Psi$ and $\Psi^\dagger$ are Bose fields possessing canonical
equal-time commutation relations, $c>0$ is a coupling constant and
$h>0$ the chemical potential. The results obtained in \cite{KozMS10}
are relative to the thermodynamic limit $L\to\infty$ of this model.

The  density operator $j(x)=\Psi^\dagger(x)\,\Psi(x)$ defines the operator of the number of particles in the interval $[0,x]$
 \begin{equation}\label{8-Q1}
 {\cal  Q}_x=\int\limits_0^xj(z)\,d z.
 \end{equation}
The generating function for the density-density correlations reads
 \begin{equation}\label{def-Gf}
 \langle e^{2\pi i\alpha {\cal Q}_x}\rangle_T=  \lim_{L\to \infty} \frac{\langle\Omega_T|e^{2\pi i\alpha {\cal Q}_x}|\Omega_T\rangle}
 {\langle\Omega_T|\Omega_T\rangle},
 \end{equation}
were $T$ is the temperature, $\alpha$ a complex number and
$|\Omega_T\rangle$ any eigenstate of $H$ that goes to the state of thermal equilibrium in the infinite volume limit.
Indeed, the correlation function of densities $\langle j(x)j(0)\rangle_T$ can be
obtained from \eqref{def-Gf} as
 \begin{equation}\label{8-cor-fun}
 \langle j(x)j(0)\rangle_T =\left.-\frac{1}{8\pi^2}\frac{\partial^2}{\partial x^2}
 \frac{\partial^2}{\partial \alpha^2}\langle e^{2\pi i\alpha {\cal Q}_x}\rangle_T \right|_{\alpha=0}.
 \end{equation}

We have shown in \cite{KozMS10} that the large-$x$ asymptotic expansion of the
generating function \eqref{def-Gf} (and respectively of the two-point
function \eqref{8-cor-fun}) is given in terms of solutions to a set
of non-linear integral equations closely related to ones arising in
the quantum transfer matrix approach
\cite{SeeBGK07,KluB90,KluBP91,Klu93,KluWZ93,Klu04}. Below, we solve
these equations in the low-temperature limit. This
computation allows us to reach two goals.

On the one hand, one can argue that the Luttinger liquid
\cite{Hal81} and conformal field theory (CFT) approaches
\cite{BelPZ84,Car84,dVegW85,Car86,BloCN86,Aff86,BogIR86,Car96}
can be used to predict the large-$x$ asymptotic behavior of the
low-temperature correlation functions in massless one dimensional
quantum models. The QNLS model does belong to this class. Thus, our
results give us a possibility to confirm  these predictions by a 
direct derivation based on the algebraic Bethe ansatz. Namely, we
show that in the low-temperature limit ($T\to 0$) the asymptotic
expansion ($x\to \infty$, $xT\to \infty$) of the density-density
correlation function takes the following form (at leading order for each oscillatory term):
 \begin{equation}\label{CFT-SinhTerm}
 \langle j(x)j(0)\rangle_T =
 D^2 \; - \frac{ (T \mathcal{Z}/v_0)^2 }{ 2 \sinh^{2}(\pi T x / v_0) } \;  + \;  \sum_{\ell \in \mathbb{Z}^* }
 A_{\ell}\; e^{2ix \ell k_{{}_F}}
 \left( \frac{  \pi T/v_0 }{  \sinh (\pi T x/v_0) }    \right)^{2 \ell^2 \mathcal{Z}^2  }
.
\end{equation}
Above appear several constants that will be defined in section~\ref{S-T0}, namely,  the average density of the gas  $D$, the Fermi momentum
$k_{{}_F}$, the velocity of sound on the Fermi boundary $v_0$ and the value of the dressed charge at Fermi boundary $\mathcal{Z}$.
The coefficients $A_{\ell}$ do not depend on $T$. All the dependence
of the amplitudes on $T$ has been gathered in the pre-factor $ ( \pi
T / v_0)^{2\ell^2 \mathcal{Z}^2}$.

This form is in full agreement with the CFT predictions. Moreover it
provides one with an asymptotic behavior that is also valid in the full
scaling region of $xT$ and in particular in the $T=0$ case.

On the other hand our approach allows us to calculate the constant
coefficients $A_{\ell}$ in \eqref{CFT-SinhTerm}. We show that
$A_{\ell}$ are related to the amplitudes of the so-called critical form
factors introduced in \cite{KitKMST10a} and arising in the study of
the model at $T=0$. More precisely, the coefficients $A_{\ell}(\pi
T/v_0)^{2 \ell^2 \mathcal{Z}^2}$ determined for the system in the
thermodynamic limit and at small but finite temperature $T$ are
equal to the amplitudes of the critical form factors corresponding
to umklapp-type excited states of momentum $2k_{{}_F} \ell$ and
determined for the system of large but finite size $L$ at zero
temperature, with the identification $v_0/T\mapsto iL$. We will show
this coincidence by means of straightforward calculations carried in
the core of this paper.

This article is organized as follows. In section~\ref{S-Results} we
recall the results obtained in \cite{KozMS10}. In section~\ref{S-T0}
we present the thermodynamics of the QNLS model at low temperature.
In particular we solve the non-linear integral equation determining
the asymptotic expansion of the correlation function in the
low-temperature approximation. This allows us to obtain the rates of
exponential decays in section~\ref{S-CL} and the constant amplitudes
in section~\ref{S-CA}. The expansion \eqref{CFT-SinhTerm} is derived
in section~\ref{S-FR}. Various estimates of the low-temperature
behavior of the integrals that we deal with are gathered in three
appendices.

\section{Long distance asymptotic behavior at general temperature\label{S-Results}}

\begin{figure}[h]
\begin{center}
 \begin{pspicture}(10,5)
 \psline[linestyle=dashed, dash=5pt 2pt]{->}(1.2,2)(9.8,2)
 \pscurve(2.4,2)(2.5,2)(3,2)(3.6,2.3)(3.8,3.1)(3.9,2.5)(3.6,2)(3.3,1.15)(3.3,0.15)(3.4,0.01)(3.6,0)
 (3.7,0.15)(3.7,0.2)(3.5,0.6)(3.5,0.7)(3.7,1)(4.1,1.3)(4.2,1.8)(4.3,1.9)(4.5,2)(5.5,2)(7.1,2)(7.5,2)
 (7.7,2.5)(7.8,3)(8.0,3.5)(8.2,3.5)(8.3,3.0)(8.2,2.5)(8.0,2.1)(7.7,1.7)(7.9,1.6)(8.1,1.7)(8.2,1.9)
 (8.3,2.0)(8.4,2.0)(8.5,2.0)(8.9,2.0)
 %
  \rput(3.9,2.2){$\scriptstyle\times$}
  \psdots(3.8,2.7) \rput(3.1,2.6){$\scriptstyle\circ$}
  \rput(3.7,3.3){$\scriptstyle\times$} \rput(2.9,3.2){$\scriptstyle\circ$}
  \rput(3.5,3.9){$\scriptstyle\times$} \rput(2.7,3.8){$\scriptstyle\circ$}
  \psdots(3.9,1.8)
  \psdots(3.8,1.3) \rput(3.1,1.4){$\scriptstyle\circ$}
  \rput(3.7,0.7){$\scriptstyle\times$} \rput(2.9,0.8){$\scriptstyle\circ$}
  \psdots(3.5,0.1) \rput(2.7,0.2){$\scriptstyle\circ$}
  \psdots(7.9,2.2)
  \psdots(8.0,2.7) \rput(8.6,2.6){$\scriptstyle\circ$}
  \psdots(8.1,3.3) \rput(8.8,3.2){$\scriptstyle\circ$}
  \rput(8.3,3.9){$\scriptstyle\times$} \rput(9.0,3.8){$\scriptstyle\circ$}
  \psdots(7.9,1.8)
  \rput(8.0,1.3){$\scriptstyle\times$} \rput(8.6,1.4){$\scriptstyle\circ$}
  \rput(8.1,0.7){$\scriptstyle\times$} \rput(8.8,0.8){$\scriptstyle\circ$}
  \rput(8.3,0.1){$\scriptstyle\times$} \rput(9.0,0.2){$\scriptstyle\circ$}
  \rput(4.5,1.4){$\hat{\cal C}_i$}
 \rput(9.2,2.2){$\mathbb{R}$}
 \psline[linewidth=2pt]{<-}(3.89,1.13)(3.82,1.08)
 \psline[linewidth=2pt]{<-}(7.86,3.19)(7.81,3.05)
\end{pspicture}
\caption{\label{hG1212} The roots $\{\Rpm\}$ are depicted by $(\circ)$,
the roots $\{\hqpm\}_i$ are depicted by $(\bullet)$. Other roots of the
equation $1+e^{-u_i(\hqpm)/T}=0$ are depicted by $(\times)$. The
contour $\hat{\cal C}_i$ bypasses the roots $\{\hqp\}_i$ from above
and the roots $\{\hqm\}_i$ from below. It also separates the points
$\{\hqpm\}_i$ from the other points $\{\hqpm\}$ as well as from all the poles of the Fermi weight $\{\Rpm\}$.
}
\end{center}
\end{figure}
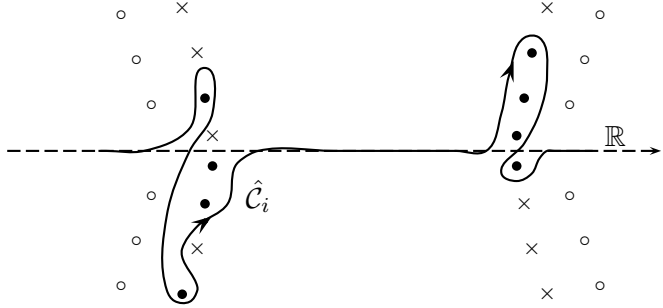

The state of the thermal equilibrium in the QNLS model is described by the
Yang--Yang equation \cite{YanY69} for the excitation energy
$\varepsilon(\lambda)$
 \begin{equation}\label{YY-orig}
 \varepsilon(\lambda)=\lambda^2-h-\frac{T}{2\pi}\int\limits_{\mathbb{R}}K(\lambda-\mu)
 \log\left(1+e^{-\frac{\varepsilon(\mu)}T}\right)\,d\mu,
 \end{equation}
and the integral equation for the total density $\rho_t(\lambda)$
 \begin{equation}\label{Inteq-Dtot}
 \rho_t(\lambda)-\frac{1}{2\pi}\int\limits_{\mathbb{R}}K(\lambda-\mu)
 \vartheta(\mu)\rho_t(\mu)\,d\mu=\frac{1}{2\pi}.
 \end{equation}
The kernel $K(\lambda)$ and the Fermi weight
$\vartheta(\lambda)$ appearing above read
 \begin{equation}\label{K-th}
 K(\lambda)=\frac{2c}{\lambda^2+c^2},\qquad
 \vartheta(\lambda)=\left(1+e^{\frac{\varepsilon(\lambda)}T}\right)^{-1}.
 \end{equation}
Below the poles of the Fermi weight will play an important role. Therefore we
introduce
the roots $\{\Rpm\}$ of the equation $1+e^{-\varepsilon(\Rpm_j)/T}=0$, where $\Rp_j$ (resp.  $\Rm_j$) belong to the upper
(resp. lower) half-plane (see Fig.~\ref{hG1212}).

The asymptotic expansion of the generating function $\langle e^{2\pi
i\alpha {\cal Q}_x}\rangle_T$ is given in terms of the solutions to
 non-linear integral equations similar to \eqref{YY-orig}. Let us
choose $n$ points ($n=0,1,\dots$) $\hqp_j$ in the upper half-plane
and $n$ points  $\hqm_j$ in the lower half-plane. We then introduce
non-linear integral equation for a function $u(\lambda)$
 \begin{equation}\label{Inteq-u1}
 u(\lambda)=\lambda^2-h_\alpha-\frac{T}{2\pi}\int\limits_{\mathbb{R}}K(\lambda-\mu)
 \log\left(1+e^{-\frac{u(\mu)}T}\right)\,d\mu+iT\sum_{j=1}^n\bigl(\theta(\lambda-\hqp_j)-
 \theta(\lambda-\hqm_j)\bigr),
 \end{equation}
where $h_\alpha=h+2\pi i\alpha T$ and
 \begin{equation}\label{theta}
 \theta(\lambda)=i\log\left(\frac{ic+\lambda}{ic-\lambda}\right),\qquad
 \theta'(\lambda)=K(\lambda).
 \end{equation}
Below we will show that, in the low-temperature limit and for
$\{\hqpm\}$ fixed, the solution to the equation \eqref{Inteq-u1}
always exists. Clearly, this solution depends on the parameters
$\{\hqpm\}$: $u(\lambda)=u(\lambda|\{\hqp\},\{\hqm\})$. By imposing
the constraints
 \begin{equation}\label{sys-s}
 1+\exp\left(-\frac{u(\hqpm_j|\{\hqp\},\{\hqm\})}T\right)=0,\qquad
 j=1,\dots,n,
 \end{equation}
we obtain a system of equations, which fixes the sets $\{\hqpm\}_i$
that are relevant for the description of the long-distance
asymptotics. The subscript $i$ enumerates these sets. The
long-distance asymptotic expansion for the generating function
$\moy{e^{2\pi i\alpha {{\cal Q}_x}}}$ can then be organized into a
sum parameterized by the functions $u_i(\lambda) \equiv
u_i(\lambda|\{\hqp\}_i,\{\hqm\}_i)$ that solve \eqref{Inteq-u1} with
a corresponding set of roots $\{\hqpm\}_i$:
 \begin{equation}\label{fin-answ}
 \moy{e^{2\pi i\alpha {{\cal Q}_x}}}=
    \sum_{i}e^{-xp[u_i]}B[u_i]
    +o\left(e^{-x\max\Re(p[u_i])}\right),
 \end{equation}
where $p[u_i]$ and $B[u_i]$ are functionals
of $u_i(\lambda)$ whose explicit form will be specified later.

Observe that equation \eqref{Inteq-u1} can be recast in the
form
 \begin{equation}\label{inteq-U1}
 u_i(\lambda)=\lambda^2-h_\alpha -\frac{T}{2\pi}\int\limits_{\hat{\cal C}_i}
 K(\lambda-\mu)\log\left(1+e^{-\frac{u_i(\mu)}T}\right)\,d\mu ,
 \end{equation}
where the contour $\hat{\cal C}_i$ is such that the roots
$\{\hqpm\}$ are located between the real axis and $\hat{\cal C}_i$. We also
demand that the contours
$\hat{\cal C}_i$ separate the  sets $\{\hqpm\}_i$  from all over possible
roots of the equation $1+e^{-{u_i(\lambda)}/T}=0$ and from all the
roots $\{\Rpm\}$  (see
Fig.~\ref{hG1212}). Then one can  interpret the asymptotic
expansion of $ \moy{e^{2\pi i\alpha {{\cal Q}_x}}}$ as being given by the
sum over the different possible choices of contours $\hat{\cal C}_i$.

%
%

The explicit expressions for the rates of
exponential decay $p[u_i]$ and the amplitudes $B[u_i]$ can be written down in terms of
integrals over the contours $\hat{\cal C}_i$.  Let us  introduce an
auxiliary functions $z_i(\lambda)$ by
 \begin{equation}\label{sol-Z}
 z_i(\lambda)=-\frac1{2\pi i}\log\left(\frac{1+
 e^{-\frac{u_i(\lambda)}T} }{
 1+ e^{-\frac{\varepsilon(\lambda)}T}} \right).
 \end{equation}
Then the rates of exponential decays take the very simple form
 \begin{equation}\label{p-decay}
 p[u_k]=i\int\limits_{\hat{\cal C}_k}z_k(\lambda)d\lambda.
 \end{equation}

The expressions for the amplitudes $B[u_i]$  are much more cumbersome. We
present them in the form
 \begin{equation}\label{formB}
 B[u_i]=B_d[u_i]B_s[u_i],\quad\mbox{where}\quad
 B_d[u_i]=\exp\Biggl(~\int\limits_{\hat{\cal C}_i}
 \frac{z_i(\lambda)z_i(\mu)}{(\lambda-\mu_+)^2}\,d\lambda\,d\mu
 \Biggr),
 \end{equation}
and the symbol $\mu_+$ means that the variable $\mu$ is slightly
shifted to the left of the oriented integration contour $\hat{\cal C}_i$.
The reason we separate the coefficients $B[u_i]$ into two
factors $B_d[u_i]$ and $B_s[u_i]$ is that these go, in the low-temperature
limit,  to the discrete and to the smooth parts of critical form
factors respectively, as defined in \cite{KitKMST10a}.

In order to describe $B_s[u_i]$ we first introduce the Cauchy transform operator
$L_{\hat{\cal C}_i}$ on the contour
$\hat{\cal C}_i$
 \begin{equation}\label{Cauchy-Tr}
 L_{\hat{\cal C}_i}[z_i](\omega)=\int\limits_{{\hat{\cal C}_i}}
 \frac{z_i(\lambda)}{\lambda-\omega}\,d\lambda,
 \end{equation}
and a functional $C_0=C_0[z_i]$
 \begin{equation}\label{7-C0}
 C_0[z_i]=\int\limits_{{\hat{\cal C}_i}}
 \frac{z_i(\lambda)z_i(\mu)}{(\lambda-\mu-i c)^2}\,d\lambda\,
 d\mu\, .
 \end{equation}
Then
 \begin{multline}\label{C2-W}
 B_s[u_i]=(e^{2\pi i\alpha}-1)^2 e^{ -C_0[z_i]}
 \frac{\det\left(I+\frac1{2\pi i}\hat U^{(1)}[z_i]\right)
 \det\left(I+\frac1{2\pi i} \hat U^{(2)}[z_i]\right)}
 {\det\left[I-{\textstyle\frac1{2\pi}}K^{(\varepsilon)}\right]
 \det\left[I-{\textstyle\frac1{2\pi}}K^{(u_i)}\right]}\numa{35}
 \times\left[e^{ L_{\hat{\cal C}_i}[z_i](\theta_1+i c)}  -e^{2\pi
i\alpha+L_{\hat{\cal C}_i}[z_i](\theta_1-i c)}\right]^{-1}
 \left[e^{ -L_{\hat{\cal C}_i}[z_i](\theta_2-i c)} -e^{2\pi i\alpha-L_{\hat{\cal C}_i}[z_i](\theta_2+i c)}\right]^{-1}.
 \end{multline}
The first line of \eqref{C2-W} contains a ratio of  Fredholm
determinants. The integral operators
$I-{\textstyle\frac1{2\pi}}K^{(\varepsilon)}$ and
$I-{\textstyle\frac1{2\pi}}K^{(u_i)}$  have the kernels
 \begin{equation}\label{Ku}
    K^{(\varepsilon)}(\lambda,\mu)=\frac{K(\lambda-\mu)}{1+e^{\frac{\varepsilon(\mu)}T}}, \qquad
    K^{(u_i)}(\lambda,\mu)=\frac{K(\lambda-\mu)}{1+e^{\frac{u_i(\mu)}T}}.
 \end{equation}
The first of these integral operators acts on the real axis and the second one acts on the contour
$\hat{\cal C}_i$. The operators $I+\frac1{2\pi i}\hat U^{(1)}[z_i]$
and $I+\frac1{2\pi i}\hat U^{(2)}[z_i]$ both act on a anticlockwise
oriented closed contour surrounding  $\hat{\cal C}_i$. Their kernels
are given by
 \begin{equation}\label{C2-U1}
 \hat U^{(1)}(w,w',[z_i])=-e^{ L_{\hat{\cal C}_i}[z_i](w)}\cdot
 \frac{K_\alpha(w-w')-K_\alpha(\theta_1-w')}{
 e^{ L_{\hat{\cal C}_i}[z_i](w+i c)}  -e^{2\pi i\alpha+L_{\hat{\cal C}_i}[z_i](w-i c)}},
 \end{equation}
and
 \begin{equation}\label{C2-U2}
 \hat U^{(2)}(w,w',[z_i])=e^{ -L_{\hat{\cal C}_i}[z_i](w')}\cdot
 \frac{K_\alpha(w-w')-K_\alpha(w-\theta_2)}{
 e^{ -L_{\hat{\cal C}_i}[z_i](w'-i c)}  -e^{2\pi i\alpha-L_{\hat{\cal C}_i}[z_i](w'+i
 c)}},
 \end{equation}
where
 \begin{equation}\label{Kalpha}
  K_\alpha(\lambda)=\frac1{\lambda+ic}-\frac{e^{2\pi i\alpha}}{\lambda-ic} \; .
 \end{equation}
Finally, observe that the kernels $\hat U^{(1,2)}(w,w',[z_i])$ as
well as the coefficient in the second line of \eqref{C2-W} depend on
arbitrary complex numbers $\theta_1$ and $\theta_2$ located  inside of
the contour where the operators $\hat U^{(1,2)}(w,w',[z_i])$ act.
One can prove (see \cite{KitKMST09b}) that the total combination
\eqref{C2-W} does not depend on the specific choice of these
parameters.



\section{Thermodynamics at low temperature\label{S-T0}}

In this section we provide a list of necessary formulae describing the
thermodynamics of the QNLS model at low temperature. We begin our discussion with
the $T=0$ case.

\subsection{Zero temperature\label{S-T00}}

It is known \cite{YanY69} that for a positive chemical potential
$h>0$ the solution $\varepsilon(\lambda)$ to the equation
\eqref{YY-orig} has two roots $\pm q(T)$ on the real axis:
$\varepsilon(\pm q(T))=0$. Hereby $\varepsilon(\lambda)>0$ for
$|\lambda|>q(T)$ and $\varepsilon(\lambda)<0$ for $|\lambda|<q(T)$.
Let $\varepsilon(\lambda)\to\varepsilon_0(\lambda)$ and $q(T)\to q$
as $T\to 0$. Then,
 \begin{equation}\label{routh-est}
\lim_{T \to 0} T\log\left(1+e^{-\frac{\varepsilon(\lambda)}T}\right)=
 \left\{\begin{array}{cc}
 {\dis 0,}&{\dis \qquad|\lambda|>q,}\num
 {\dis -\varepsilon_0(\lambda),}&{\dis \qquad|\lambda|<q,}
 \end{array}\right.  \; .
 \end{equation}
It is then straightforward to check that equation \eqref{YY-orig} turns into a linear integral equation for the
dressed energy $\varepsilon_0(\lambda)$:
 \begin{equation}\label{YY-0}
 \varepsilon_0(\lambda)-\frac{1}{2\pi}\int\limits_{-q}^qK(\lambda-\mu)
 \varepsilon_0(\mu)\,d\mu=\lambda^2-h, \qquad \varepsilon_0(\pm
 q)=0.
 \end{equation}

At $T=0$ the state of  thermal equilibrium goes to the ground
state of the QNLS model. The Fermi weight $\vartheta(\lambda)$
\eqref{K-th} turns into the characteristic function of the interval
$[-q,q]$. Therefore the equation \eqref{Inteq-Dtot} for the density
takes the form
 \begin{equation}\label{Inteq-DtotT0}
 \rho_t(\lambda)-\frac{1}{2\pi}\int\limits_{-q}^qK(\lambda-\mu)
 \rho_t(\mu)\,d\mu=\frac{1}{2\pi},\qquad T=0.
 \end{equation}

Another important characteristic of the ground state is the dressed
charge $Z(\lambda)$. In the QNLS model it is proportional to the
density $Z(\lambda)=2\pi \rho_t(\lambda)$. Below, we will use a
special notation for the value of the dressed charge on the Fermi boundary ${\cal
Z}=Z(\pm q)$. A formal expression for ${\cal Z}$ can be given in
terms of the resolvent to the operator $I-\frac1{2\pi}K$ acting on
the interval $[-q,q]$

 \begin{equation}\label{res-dc}
 {\cal Z}=1+\int\limits_{-q}^qR(\lambda,\pm q)\,d\lambda,
 \end{equation}
where
 \begin{equation}\label{Resolv}
 R(\lambda,\xi)-\frac{1}{2\pi}\int\limits_{-q}^qK(\lambda-\mu)
 R(\mu,\xi)\,d\mu=\frac{1}{2\pi}K(\lambda-\xi).
 \end{equation}

Finally, we give the formulae for the constants appearing in \eqref{CFT-SinhTerm}, namely the average density $D$, the Fermi
momentum $k_{{}_F}$, and the velocity of the sound on the Fermi
boundary $v_0$:
 \begin{equation}\label{v0}
 D=\int\limits_{-q}^q\rho_t(\lambda)\,d\lambda, \qquad k_{{}_F}=\pi D,\qquad v_0=\frac{\varepsilon'_0}{\cal
 Z},
 \end{equation}
where we denoted $\varepsilon'_0\equiv\varepsilon'_0(q)$.

\subsection{The dressed energy at low temperature}

We now describe the power-law corrections to the Yang--Yang equation
\eqref{YY-orig}, which appear at low but non-zero temperature.  For
$T$ small enough the solution to the equation \eqref{YY-orig}
$\varepsilon(\lambda)$ has the following expansion
\cite{JohM72,MezN92,Sla99}:
 \begin{equation}\label{eps-dec}
 \varepsilon(\lambda)=\sum_{k=0}^2T^k\varepsilon_k(\lambda)+O(T^3).
 \end{equation}
The functions $\varepsilon_1(\lambda)$ and $\varepsilon_2(\lambda)$
can be found from the analysis of the integral in \eqref{YY-orig} in
small vicinities of the Fermi boundaries $\pm q$. The details of
this analysis are given in appendix~\ref{SM-Int} (see also
\cite{JohM72,MezN92,Sla99}).

Using \eqref{Intgr} and substituting
the expansion \eqref{eps-dec} into the Yang--Yang equation we obtain
 \begin{multline}\label{YY-T-small}
 \sum_{k=0}^2 T^k\varepsilon_k(\lambda)-\frac{1}{2\pi} \sum_{k=0}^2 T^k\int\limits_{-q}^qK(\lambda-\mu)
 \varepsilon_k(\mu)\,d\mu=\lambda^2-h\\
 -\frac{T^2\pi}{12\varepsilon'_0}\bigl(
 K(\lambda-q)+K(\lambda+q)\bigr)
 -\frac{T^2K(\lambda-q)\varepsilon_1^2(q)}{4\pi\varepsilon'_0}-\frac{T^2K(\lambda+q)\varepsilon_1^2(-q)}{4\pi\varepsilon'_0}
 +O(T^3).
 \end{multline}
What follows from this analysis is that $\varepsilon_1(\lambda)=0$ and
 \begin{equation}\label{e12}
 \varepsilon_2(\lambda)=-\frac{\pi^2}{6\varepsilon'_0}\bigl(R(\lambda,q)+R(\lambda,-q)\bigr),
 \end{equation}
where $R(\lambda,\mu)$ corresponds to the resolvent defined by \eqref{Resolv}.

\subsection{The poles of the Fermi weight}

We now consider the low-temperature behavior of the roots $\Rpm_k$ to
the equation $1+e^{-\frac{\varepsilon(\lambda)}T}=0$, namely the solutions to
 $\varepsilon(\lambda)=2\pi iT(k+1/2)$, $k\in
\mathbb{Z}$. Obviously all these roots collapse to $q$ or $-q$ when
$T\to 0$. Therefore setting
$\Rpm_k=q +T\TRpm_k+O(T^2)$ (resp. $\Rpm_k=-q +T\TRpm_k+O(T^2)$) and substituting
these expansions into
$\varepsilon(\Rpm_k)=2\pi iT(k+1/2)$ we find
 \begin{equation}\label{equation-rk}
 \begin{array}{l}
 \varepsilon\bigl( q +T\TRpm_k+O(T^2)\bigr)= T
 \TRpm_k\varepsilon'_0+O(T^2)= 2\pi iT(k+1/2),\\
 \varepsilon\bigl(- q +T\TRpm_k+O(T^2)\bigr)= -T
 \TRpm_k\varepsilon'_0+O(T^2)= 2\pi iT(k+1/2).
 \end{array}
 \end{equation}
Thus, in the linear approximation in $T$, we obtain two series of
roots
 \begin{equation}\label{2-ser-rk}
 \left\{
 \begin{array}{c}
 {\dis \Rp_k=\pm q+\frac{2\pi iT}{\varepsilon'_0}(k+1/2)+O(T^2), \quad k\ge
 0,}\num
  {\dis \Rm_k=\pm q+\frac{2\pi iT}{\varepsilon'_0}(k+1/2)+O(T^2), \quad k< 0.}
  \end{array}\right.
  \end{equation}
We will refer to the roots collapsing to $+q$ as the right series
and the roots collapsing to $-q$ as the left series.

\subsection{Low-temperature limit of the $u(\lambda)$ integral  equation\label{Int-EQ}}

From now on,  we focus on a fixed contour $\hat{\cal C}_i$ and
consider the associated contribution to the asymptotic behavior of
the generating function. Therefore, below, we will omit the
subscript $i$ in the notations of the contour $\hat{\cal C}_i$ and
of the functions $u_i(\lambda)$, $z_i(\lambda)$, \textit{etc}
associated with it.

Let $\hat{\cal C}$ be the contour  bypassing $n$ points $\hqp$ in the
upper half-plane from above and $n$ points $\hqm$ in the lower
half-plane from below, where $n$ is an arbitrary, but fixed
non-negative integer.  These points $\hqpm$ are roots
of the equation $1+e^{-u(\hqpm)/T}=0$. It is important for our purpose to fix
the limits of these roots at $T=0$.

The $T \to 0$ limit of equation \eqref{Inteq-u1} coincides with the
one of the Yang--Yang equation,  hence $\left.u(\lambda)\right|_{T=0}=\varepsilon_0(\lambda)$.
Therefore it is reasonable to expect that, similarly to the points $\Rpm_k$,
the roots $\hqpm_k$ collapse to $q$ or to $-q$ in the  $T \to 0$ limit. Thus, in the
low-temperature limit these roots should form two series. There will be
\begin{itemize}
\item $n^+_p$ roots $\hqp$  and $n^+_h$ roots $\hqm$
belonging to the right series;
\item  $n^-_p$ roots $\hqp$ and $n^-_h$ roots $\hqm$ belonging to the
left series.
\end{itemize}

\noindent  Obviously, there exists an integer $\ell$, $-n\le \ell\le n$, such that the numbers $n_p^{\pm}$ and $n_h^{\pm}$ are related by
 \begin{equation}\label{sum-n}
 n^+_p+n^-_p=n^+_h+n^-_h=n,\qquad
 n^+_p-n^+_h=n^-_h-n^-_p=\ell \; .
 \end{equation}
Therefore, for $T$ small enough, one deals with the following structure for the distribution of roots $\hqpm$:
 \begin{align}\label{s+}
 \{\hqp\}&=\{q+iT\hat{\eta}^+_k\}_{n^+_p}\cup\{-q+iT\hat{\eta}^-_k\}_{n^-_p},\qquad
 \Re( \hat{\eta}^\pm_k)>0,\\
 \{\hqm\}&=\{q-iT\hat{\xi}^+_k\}_{n^+_h}\cup\{-q-iT\hat{\xi}^-_k\}_{n^-_h},\qquad
 \Re( \hat{\xi}^\pm_k)>0.\label{s-}
 \end{align}
The parameters $\hat{\eta}^\pm_k$ and $\hat{\xi}^\pm_k$ admit the Taylor expansions $\hat{\eta}^\pm_k={\eta}^{\pm}_k + \text{O}(T)$ and $\hat{\xi}^\pm_k={\xi}^{\pm}_k + \text{O}(T)$. They appear in the non-linear integral equation \eqref{Inteq-u1} defining $u(\lambda)$
and should be computed by solving  the conditions
 \begin{equation}\label{def-roots}
 \exp\left(-\frac{u(\pm q+iT\hat{\eta}^\pm_k)}T\right)=\exp\left(-\frac{u(\pm q-iT\hat{\xi}^\pm_k)}T\right)=-1.
 \end{equation}

Substituting the parameterizations \eqref{s+}, \eqref{s-} into the equation
\eqref{Inteq-u1} and expanding up to the second order in $T$ we are led to
 \begin{equation}\label{Int-eq-orig}
 u(\lambda)=\lambda^2-h-\frac{T}{2\pi}\int\limits_{\mathbb{R}}K(\lambda-\mu)\log\left(1+e^{-\frac{u(\mu)}T}\right)\,d\mu
 +TG_1(\lambda)+T^2G_2(\lambda)+O(T^3).
 \end{equation}
Here
 \begin{equation}\label{g1}
 G_1(\lambda)=-2\pi i\alpha-
 i\ell\int\limits_{-q}^qK(\lambda-\mu)\,d\mu,
 \end{equation}
and
 \begin{equation}\label{g2}
 G_2(\lambda)=K(\lambda-q)\left(\sum_{j=1}^{n^+_p}\eta^+_j+\sum_{j=1}^{n^+_h}\xi^+_j\right)
 +K(\lambda+q)\left(\sum_{j=1}^{n^-_p}\eta^-_j+\sum_{j=1}^{n^-_h}\xi^-_j\right).
 \end{equation}
It is natural to expect that the solution to \eqref{Int-eq-orig} has a form similar to \eqref{eps-dec}
 \begin{equation}\label{u-dec}
 u(\lambda)=\sum_{k=0}^2T^ku_k(\lambda)+O(T^3),
 \end{equation}
where, as we have already argued, $u_0(\lambda)=\varepsilon_0(\lambda)$.  Substituting
\eqref{u-dec} into \eqref{Int-eq-orig} and using \eqref{Intgr-u}, we
obtain linear integral equations satisfied by the unknown functions $u_1(\lambda)$ and
$u_2(\lambda)$:
 \begin{equation}\label{Ieq-u1}
 u_1(\lambda)-\frac{1}{2\pi} \int\limits_{-q}^qK(\lambda-\mu)
 u_1(\mu)\,d\mu=G_1(\lambda),
 \end{equation}
 \begin{multline}\label{Ieq-u2}
 u_2(\lambda)-\frac{1}{2\pi} \int\limits_{-q}^qK(\lambda-\mu)
 u_2(\mu)\,d\mu=G_2(\lambda)-\frac{\pi}{12\varepsilon'_0}\bigl(
 K(\lambda-q)+K(\lambda+q)\bigr)\\
 -\frac{K(\lambda-q)u_1^2(q)}{4\pi\varepsilon'_0}-\frac{K(\lambda+q)u_1^2(-q)}{4\pi\varepsilon'_0}.
 \end{multline}
It is then easy to see that the function $u_1(\lambda)-2\pi i\ell$
satisfies (up to a multiplicative factor) the equation
\eqref{Inteq-DtotT0} for the total density at $T=0$. As we have
already mentioned, on has that $2\pi\rho_t(\lambda)=Z(\lambda)$  in
the case of the QNLS model, with $Z(\lambda)$ being the dressed
charge. Hence,
 \begin{equation}\label{u1}
 u_1(\lambda)=u_1(-\lambda)=-2\pi i\alpha_\ell Z(\lambda)+2\pi i\ell, \qquad
 \alpha_\ell =\alpha+ \ell.
 \end{equation}

The solution to equation \eqref{Ieq-u2} can be expressed in terms
of the resolvent $R(\lambda,\mu)$ \eqref{Resolv}
 \begin{multline}\label{u2}
 u_2(\lambda)=R(\lambda,q)\left[2\pi\sum_{j=1}^{n^+_p}\eta^+_j+2\pi\sum_{j=1}^{n^+_h}\xi^+_j-
 \frac1{2\varepsilon'_0}\left(\frac{\pi^2}3+u_1^2(q)\right)\right]\\
 +R(\lambda,-q)\left[2\pi\sum_{j=1}^{n^-_p}\eta^-_j+2\pi\sum_{j=1}^{n^-_h}\xi^-_j-
 \frac1{2\varepsilon'_0}\left(\frac{\pi^2}3+u_1^2(q)\right)\right].
 \end{multline}

It remains to fix the leading Taylor coefficients $\eta^\pm_k$ and
$\xi^\pm_k$. These can be parameterized by sets of integers, exactly
as it was the case for the roots $\Rpm_k$ \eqref{2-ser-rk}. More
precisely, one has
  \begin{equation}\label{eq-rrots}
  \begin{array}{c}
  u(\pm q+iT\hat{\eta}^\pm_k)=\pm 2\pi iT(p^\pm_k-\frac12),
  \\
    u(\pm q-iT\hat{\xi}^\pm_k)=\mp 2\pi iT(h^\pm_k-\frac12), 
  \end{array}
  \end{equation}
where $p^\pm_k$ and $h^\pm_k$ are  integers. As a consequence, in
the linear order in $T$, we obtain
  \begin{equation}\label{sol-rrots}
  \begin{array}{c}
  \varepsilon'_0\eta^\pm_k=2\pi(p^\pm_k-\frac12)\pm iu_1(q),\\
  \varepsilon'_0\xi^\pm_k= 2\pi (h^\pm_k-\frac12) \mp iu_1(q),
  \end{array}
  \end{equation}
where $u_1(\lambda)$ is given by \eqref{u1}.

{\sl Remark.} Let $u_1\equiv u_1(\pm q)=2\pi i(\ell-\alpha_\ell{\cal
Z})$. From now on we assume that $u_1$ satisfies the constraint
 \begin{equation}\label{Constr}
 -\pi<\Im(u_1)<\pi.
 \end{equation}
Note that the generating function \eqref{def-Gf} is periodic over
$\alpha$ \cite{KitKMST07}: $\langle e^{2\pi i\alpha {\cal
Q}_x}\rangle_T= \langle e^{2\pi i(\alpha+1) {\cal Q}_x}\rangle_T$.
Due to this property  the condition above always can be satisfied by
appropriate choice of the parameter $\alpha$. Therefore the
constraint \eqref{Constr} does not imply any additional restrictions
for the parameters of the model.

We stress that the condition \eqref{Constr} is purely technical. It
allows us to simplify some intermediate calculations. In particular,
it follows from \eqref{Constr} that all the integers $p^\pm_k$ and
$h^\pm_k$ in \eqref{sol-rrots} are positive. However, one can
proceed further without use of the inequality \eqref{Constr}.

Thus, in this way, we have found the solution $u(\lambda)$ to the
equation \eqref{Inteq-u1} up to $O(T^2)$ terms and the roots
$\hqpm_k$ up to $O(T)$ terms. There is no fundamental obstacle to
finding higher order corrections to $u(\lambda)$ and $\hqpm_k$.
However, for our purposes, the results obtained here are already
sufficient.

\section{Correlation lengths\label{S-CL}}

In this section we compute the rate $p[u]$ of the correlation
function exponential decay.  In the case of the QNLS model, the
conformal dimensions giving rise to the critical exponents in the
asymptotic expansion \eqref{CFT-SinhTerm}  were calculated in
\cite{BerM88a,BerM88b}. We now obtain these results by taking the
$T\to 0$ limit of equation \eqref{p-decay}.

We have already shown in the work \cite{KozMS10} how the trivial
constant term in \eqref{CFT-SinhTerm} can be deduced from  our
approach to the asymptotics at finite temperature. More precisely,
this constant stems from the contribution of the contour $\hat{\cal
C}=\mathbb{R}$, in other words the case where the sets of the roots
$\{\hqpm\}$ are empty ($n=0$). Therefore, in the following, we will
only consider the case of non-empty sets $\{\hqpm\}$ (although the
results of our analysis remain valid for $n= 0$ as well).

By moving the contour $\hat {\cal C}$ to the real axis, equation \eqref{p-decay} boils down to
 \begin{equation}\label{Cor-rad-or0}
 p[u]=i\int\limits_{\mathbb{R}}z(\mu)\,d\mu-i\sum_{k=1}^{n}(\hqp_k-\hqm_k).
 \end{equation}
The integral over $\mathbb{R}$ can be estimated to the leading order in $T$ with the help of \eqref{Intgr}, \eqref{Intgr-u}.
In its turn, the finite sum is estimated directly by inserting the Taylor expansions of the roots $\hat{s}^{\pm}_k$.
Ultimately, one gets that, to the linear order in $T$,
 \begin{equation}\label{Cor-rad1}
 p[u]=-2 i\alpha_\ell k_{{}_F}-\frac{T{\cal Z}u_1^2}{2\pi
 \varepsilon'_0}+T{\cal Z}\left(\sum_{j=1}^{n^+_p}\eta^+_j+\sum_{j=1}^{n^-_p}\eta^-_j+\sum_{j=1}^{n^+_h}\xi^+_j
 +\sum_{j=1}^{n^-_h}\xi^-_j\right) +O(T^2) .
 \end{equation}
where we have used \eqref{res-dc} and \eqref{v0}. We recall  also
that $u_1=u_1(q)=2\pi i(\ell-\alpha_\ell{\cal Z})$. Finally, it
remains to use that  $\eta^\pm_k$, $\xi^\pm_k$ are  given  by
\eqref{sol-rrots}. This leads to
 \begin{equation}\label{Cor-rad2}
 p[u]=-2 i\alpha_\ell k_{{}_F}+\frac{2\pi
 T}{v_0}\left[\left(\alpha_\ell  {\cal Z}\right)^2-\ell^2-n
 +\sum_{j=1}^{n^+_p}p^+_j+\sum_{j=1}^{n^-_p}p^-_j+\sum_{j=1}^{n^+_h}h^+_j
 +\sum_{j=1}^{n^-_h}h^-_j\right] +O(T^2).
 \end{equation}
%



\section{Constant amplitude\label{S-CA}}

In this section, we compute the low-temperature limit of the
constant coefficients $B_d[u]$ \eqref{formB} and $B_s[u]$
\eqref{C2-W}. We prove that in this limit, when properly normalized
in the temperature, $B[u]$ goes to the amplitude of a critical form
factor. The latter form factors correspond to expectation values of
local operators taken between the ground state and excited states
where all rapidities of the particles and holes are located on the
Fermi boundary. We first recall several definitions and results
concerning the form factors in the QNLS model. The reader
can find a more detailed exposition in \cite{KitKMST10a}\footnote[1]{%
Formally the work \cite{KitKMST10a} deals with form factors of the
$XXZ$ spin chain, however the results obtained there can be easily
reduced to the case of the QNLS model.}.

The form factors of the QNLS model can be parameterized by the
rapidities of particles and holes
\cite{LieL63,Lie63,LieM66L,BogIK93L}. If, in the thermodynamic limit
($L\to\infty$) all the rapidities are located on the Fermi
boundaries $\pm q$, then the corresponding form factor is called
critical form factor \cite{KitKMST10a}. Hereby the distribution of
the rapidities between $+q$ and $-q$ is important.

Consider a critical form factor parameterized by the rapidities of
$n$ particles and $n$ holes. Assume that,  in the thermodynamic
limit, there is $n^+_p$ (resp. $n^+_h$) rapidities of the particles
(resp. holes) going to $+q$ and $n^-_p$ (resp. $n^-_h$) rapidities
of the particles (resp. holes) going to $-q$.
 We say that a given form factor belongs to the $\mathbf{P}_\ell$ class, if the numbers $n^\pm_{p,h}$
satisfy  the conditions gathered in \eqref{sum-n}, with $\ell$ being some fixed integer.

The critical form factors can be presented as a product of a smooth
and a discrete part (see \cite{KitKMST10a}). The smooth part has a well
defined thermodynamic limit $L\to\infty$. The discrete part,
strictly speaking, has no thermodynamic limit. First of all, it
scales to zero as some negative power of $L$, when $L\to\infty$.
Second, it not only depends on the rapidities of the particles and
holes (which are equal to $\pm q$), but also on the quantum
numbers associated with the excited state.

In the following, we show that the factor $B_d[u]$ in \eqref{formB}
exactly reproduces  the discrete  part of the critical form factor
of the $\mathbf{P}_\ell$ class, provided  the distribution
\eqref{sum-n} is fixed. Hereby the role of large $L$ is played by
the inverse temperature: $v_0/iT \leftrightarrow L$. The integers
$p^\pm_j$ and $h^\pm_j$ (see \eqref{eq-rrots}) play the role of the
quantum numbers describing particles and holes.

The coefficient $B_s[u]$ \eqref{C2-W} gives the smooth part of the
critical form factor. We first focus on the analysis related with
$B_s[u]$ as the computation of its $T\rightarrow 0$ limit is simpler
then for $B_d[u]$.

\subsection{Smooth part}

The coefficient $B_s[u]$ can be seen as mostly depending on integrals of the
following type:
 \begin{equation}\label{Int-type}
 I_f=\int\limits_{\hat{\cal C}} f'(\lambda)z(\lambda)\,d\lambda,
 \end{equation}
where $z(\lambda)$ is given by \eqref{sol-Z} and $f(\lambda)$ is
holomorphic in some  domain containing $\hat{\cal C}$ and
$\mathbb{R}$. Then moving $\hat{\cal C}$ to $\mathbb{R}$ we obtain
 \begin{equation}\label{Main-f}
 I_f\to  \int\limits_{\mathbb{R}}
  f'(\lambda)z(\lambda)\,d\lambda-\ell\bigl(f(q)-f(-q)\bigr), \qquad
  T\to 0,
  \end{equation}
since all roots  $\{\hqpm\}$ go to $\pm q$ at $T\to 0$. Using that, at $T=0$
$z(\lambda)=0$ for $|\lambda|>q$ and
$z(\lambda)=u_1(\lambda)/2\pi i$ for $|\lambda|<q$, we find
 \begin{equation}\label{Main-f2}
  \lim_{T\to 0}  \int\limits_{\mathbb{R}}
  f'(\lambda)z(\lambda)\,d\lambda= \frac{1}{2\pi i}
  \int\limits_{-q}^q
  f'(\lambda)u_1(\lambda)\,d\lambda,
  \end{equation}
and hence, due to \eqref{u1}
 \begin{equation}\label{Main-f3}
 \lim_{T\to 0}  \int\limits_{\hat{\cal C}} f'(\lambda)z(\lambda)\,d\lambda=
  -\alpha_\ell \int\limits_{-q}^q
  f'(\lambda)Z(\lambda)\,d\lambda .
  \end{equation}

Using this prescription we obtain for the limit of the Cauchy
transforms
 \begin{equation}\label{Lim-Cauchy}
 \lim_{T\to 0}  L_{\hat{\cal C}}[z](w+i \gamma c)=-\alpha_\ell L_{[-q,q]}[Z](w+i \gamma
 c),\qquad \gamma=0,\pm1.
 \end{equation}
Similarly
 \begin{equation}\label{7-C0-lim}
 \lim_{T\to 0}  C_0[z]=\alpha_\ell^2\int\limits_{-q}^{q}
 \frac{Z(\lambda)Z(\mu)}{(\lambda-\mu-i c)^2}\,d\lambda\,
 d\mu.
 \end{equation}

Another type of integrals arises in the Fredholm determinant
${\det}_{\hat{\cal
C}}\left[I-{\textstyle\frac1{2\pi}}K^{(u)}\right]$. Recall that this
operator acts on the contour $\hat{\cal C}$ as
 \begin{equation}\label{act-Ku}
 \left[I-{\textstyle\frac1{2\pi}}K^{(u)}\right]f(\lambda)=
 f(\lambda)-{\textstyle\frac1{2\pi}}\int\limits_{\hat{\cal
 C}}K^{(u)}(\lambda,\mu)f(\mu)\,d\mu,
 \end{equation}
where $K^{(u)}(\lambda,\mu)$ is given by \eqref{Ku}. If $f(\lambda)$
is holomorphic in a domain containing $\hat{\cal C}$ and
$\mathbb{R}$, then one can easily see that
 \begin{equation}\label{act-KuCR1}
 \int\limits_{\hat{\cal
 C}}K^{(u)}(\lambda,\mu)f(\mu)\,d\mu=\int\limits_{\mathbb{R}}K^{(u)}(\lambda,\mu)f(\mu)\,d\mu+O(T), \qquad T\to 0.
 \end{equation}
Since $u(\lambda)=\varepsilon_0(\lambda)$ at $T=0$ we conclude that, in the $T\to 0$ limit,
the action of the operator $I-{\textstyle\frac1{2\pi}}K^{(u)}$
coincides with the one of
$I-{\textstyle\frac1{2\pi}}K^{(\varepsilon)}$. The action of this last operator clearly reduces to the interval $[-q,q]$ when $T=0$. Thus,
 \begin{equation}\label{lim-det}
 \lim_{T\to
 0}{\det}_{\mathbb{R}}\left[I-{\textstyle\frac1{2\pi}}K^{(\varepsilon)}\right]=
 \lim_{T\to 0}{\det}_{\hat{\cal
 C}}\left[I-{\textstyle\frac1{2\pi}}K^{(u)}\right]=
 {\det}_{[-q,q]}\left[I-{\textstyle\frac1{2\pi}}K\right].
 \end{equation}

Substituting all these results into \eqref{C2-W} we immediately
reproduce the smooth part of the critical form factor obtained in
\cite{KitKMST10a}.  We give these rather cumbersome expressions
in appendix~\ref{SPoA-Sec}.

\subsection{Discrete part}

The $T\to0$ limit of the factor $B_d[u]$ \eqref{formB} is more
involved. In order to compute it, we first deform the contour
$\hat{\cal C}$ to the real axis. This provides an alternative expression for $B_d[u]$, that was
originally obtained in  \cite{Sla10a}
 \begin{multline}\label{Move-to-R}
 B_d[u]=
  \exp\biggl(~\dis\int\limits_{\mathbb{R}}
 \frac{z(\lambda)z(\mu)}{(\lambda-\mu_+)^2}\,d\lambda\,d\mu\biggr)\cdot
 \left(\det_n\frac1{\hqp_j-\hqm_k}\right)^2\\
 \times\prod_{j=1}^n e^{2L_{\mathbb{R}}[z](\hqm_j)-2L_{\mathbb{R}}[z](\hqp_j)}
 \left(\Bigl.\partial_\lambda e^{-2\pi iz(\lambda)}\Bigr|_{\lambda=\hqm_j}\right)^{-1}
 \left(\partial_\lambda e^{-2\pi iz(\lambda)}\Bigr|_{\lambda=\hqp_j} \right)^{-1} .
 \end{multline}

Consider the behavior of the Cauchy determinant in \eqref{Move-to-R}
at $T\to 0$. We have
 \begin{equation}\label{qCauchy}
 \left(\det_{n}\frac1{\hqp_j-\hqm_k}\right)^2=\frac{\prod\limits_{j>k}^{n}(\hqp_j-\hqp_k)^2
 (\hqm_j-\hqm_k)^2}{\prod\limits_{j,k=1}^{n}(\hqp_j-\hqm_k)^2}.
 \end{equation}
Now we should substitute here \eqref{s+}, \eqref{s-} and
\eqref{sol-rrots}. Hereby at $T\to 0$ we can set
$(\hqpm_j-\hqpm_k)^2=(\hqpm_j-\hqmp_k)^2=4q^2$, if the roots belong
to the different series. Then we obtain
 \begin{multline}\label{det-Cauchy}
 \lim_{T\to 0}\left(T^{n-\ell^2}\det_n\frac1{\hqp_j-\hqm_k}\right)^2=(-1)^{n+\ell}\left(\frac{q\varepsilon'_0} {\pi }\right)^{-2\ell^2}
 \left(\frac{\varepsilon'_0} {2\pi }\right)^{2n}\\
 \times \frac{\prod\limits_{j>k}^{n^+_p}(p^+_j-p^+_k)^2
 \prod\limits_{j>k}^{n^+_h}(h^+_j-h^+_k)^2}{\prod\limits_{j=1}^{n^+_p}\prod\limits_{k=1}^{n^+_h}(p^+_j+h^+_k-1)^2}
 \frac{\prod\limits_{j>k}^{n^-_p}(p^-_j-p^-_k)^2
 \prod\limits_{j>k}^{n^-_h}(h^-_j-h^-_k)^2}{\prod\limits_{j=1}^{n^-_p}\prod\limits_{k=1}^{n^-_h}(p^-_j+h^-_k-1)^2}.
 \end{multline}

Consider now the low-temperature behavior of the derivatives
$\partial_\lambda e^{-2\pi iz(\lambda)}$ at $\lambda=\hqpm_j$. We
have
 \begin{equation}\label{e-z}
 e^{-2\pi iz(\lambda)}=\frac{1+e^{-\frac{u(\lambda)}T}}
 {1+e^{-\frac{\varepsilon(\lambda)}T}}.
 \end{equation}
Since $1+e^{-\frac{u(\hqpm_j)}T}=0$ we obtain
 \begin{equation}\label{der-e-z}
 T\Bigl.\partial_\lambda e^{-2\pi iz(\lambda)}\Bigr|_{\lambda=\hqpm_j}=\frac{ - u'(\hqpm_j)}
 {1+e^{-\frac{\varepsilon(\hqpm_j)}T}}=\frac{ - u'(\hqpm_j)}
 {1-e^{-\frac{\varepsilon(\hqpm_j)-u(\hqpm_j)}T}}.
 \end{equation}
Substituting here \eqref{u-dec}, \eqref{u1} we arrive at
 \begin{equation}\label{gFp}
 \lim_{T\to 0}T\Bigl.\partial_\lambda e^{-2\pi iz(\lambda)}\Bigr|_{\lambda=\hqpm_j}=
 \left\{\begin{array}{cl}
  -  \varepsilon'_0\left(1-e^{-2\pi i\alpha_\ell {\cal Z}}\right)^{-1},&\mbox{if $\hqpm_j$ belongs to the right series,}\\
   \varepsilon'_0\left(1-e^{-2\pi i\alpha_\ell {\cal Z}}\right)^{-1},&\mbox{if $\hqpm_j$ belongs to the left series.}
 \end{array}\right.
 \end{equation}

The estimate of the $T\to 0$ behavior of the Cauchy transforms
$L_{\mathbb{R}}[z](\hqpm_j)$ is a more complicated problem. It is
easy to see that one cannot use \eqref{Lim-Cauchy} in this case.
Indeed, on the one hand the Cauchy transform $L_{[-q,q]}[z](\omega)$
on the interval $[-q,q]$ has a logarithmic singularity at
$\omega=\pm q$. On the other hand, we have seen that all $\hqpm_j$
go to $\pm q$ at $T\to 0$, therefore $L_{\mathbb{R}}[z](\hqpm_j)$
should diverge as a multiple of  $\log T $ in the low-temperature
limit. In fact, the equation \eqref{Lim-Cauchy} only allows one
 to access to this divergent part; however it does not give an access to the finite
part of the $T\to 0$ behavior of the Cauchy transform.

Similar problem occurs at studying the $T\to 0$ behavior of the
double integral
 \begin{equation}\label{Doub-intR}
 A=\int\limits_{\mathbb{R}}
 \frac{z(\lambda)z(\mu)}{(\lambda-\mu_+)^2}\,d\lambda\,d\mu.
 \end{equation}
The above double integral can be reduced to a single one
 \begin{equation}\label{Doub-intR-sing}
 A=\int\limits_{\mathbb{R}}
 \frac{z(\lambda)z(\mu)}{(\lambda-\mu_+)^2}\,d\lambda\,d\mu=
 \int\limits_{\mathbb{R}}\bigl(\partial_\mu L_{\mathbb{R}}[z](\mu_+)\bigr)\cdot z(\mu)\,d\mu.
 \end{equation}
Thus, in order to estimate this integral at $T\to 0$ one should know
the behavior of $L_{\mathbb{R}}[z](\mu)$ on the whole real axis
including the points $\pm q$. This question is studied in details in
appendix~\ref{sing-int}. Here we restrict ourselves to present the results of these computations.

First of all, we list the leading $T\to 0$ behavior of the exponents of the Cauchy transforms
$L_{\mathbb{R}}[z](\hqpm_j)$. There are four cases to distinguish:
\begin{itemize}

\item $\hqp_k$ belongs to the right or to the left series,
\item $\hqm_k$ belongs to the right or to the left series.

\end{itemize}

Using the parametrization \eqref{s+}, \eqref{s-} and
\eqref{sol-rrots} we have
 \begin{align}\label{alpha+s+}
 \lim_{T\to 0}  \bigg\{
 e^{L_{\mathbb{R}}[z](\pm q+iT\hat{\eta}^\pm_k)}
 \left(\frac{q\varepsilon'_0} {\pi T}\right)^{\pm \frac{u_1}{2\pi i} }   \bigg\}
 &=\exp\bigg\{ -\alpha_\ell \int\limits_{-q}^q\frac{Z(\mu)-{\cal Z}}{\mu\mp q}\,d\mu- \frac{u_1}{4} \bigg\}
 \frac{\Gamma(p^\pm_k)}{\Gamma(p^\pm_k \mp \frac{u_1}{2\pi i} )},\\
 \label{alpha-s+}
 \lim_{T\to 0} \bigg\{ e^{L_{\mathbb{R}}[z](\pm q-iT \hat{\xi}^\pm_k)}
 \left(\frac{q\varepsilon'_0} {\pi T}\right)^{\pm \frac{u_1}{2\pi i} } \bigg\}
 &=\exp\bigg\{-\alpha_\ell \int\limits_{-q}^q\frac{Z(\mu)-{\cal Z}}{\mu\mp q}\,d\mu+\frac{u_1}{4}\bigg\}
  \frac{\Gamma(h^\pm_k \pm \frac{u_1}{2\pi i} )}{\Gamma(h^\pm_k)}.
 \end{align}

Thus, for a given partition \eqref{s+}, \eqref{s-} of roots $\hqpm_j$
into the right and left series, we obtain
 \begin{multline}\label{prod-alpha}
 \lim_{T\to 0}\prod_{j=1}^n e^{2L_{\mathbb{R}}[z](\hqm_j)-2L_{\mathbb{R}}[z](\hqp_j)}
 \left(\frac{q\varepsilon'_0} {\pi T}\right)^{-\frac{2\ell u_1}{\pi i} }
 =\exp\left\{4 \ell \alpha_\ell
 \int\limits_{-q}^q\frac{Z(\mu)-{\cal Z}}{\mu-q}\,d\mu+2i\pi n\alpha_\ell {\cal Z}\right\}
 \\
 \times
 \Gamma^2\left(
 \begin{array}{cccc} \{p^+_k-\frac{u_1}{2\pi i} \},& \{h^+_k+\frac{u_1}{2\pi i} \}, & \{p^-_k+\frac{u_1}{2\pi i} \},&
 \{h^-_k-\frac{u_1}{2\pi i}\}\\
 \{p^+_k\},& \{h^+_k\}, & \{p^-_k\},& \{h^-_k\}
 \end{array}\right),
 \end{multline}
where we  use the standard hypergeometric type notation for ratios
of $\Gamma$-functions:
 \begin{equation}
 \Gamma\left(
 \begin{array}{c} a_1\ ,\ \dots\ ,\ a_{p}\\
                              b_1\ ,\ \dots\ ,\ b_{q}
 \end{array}\right)
  =  \prod_{k=1}^{p} \Gamma(a_k) \cdot \prod_{k=1}^{q} \Gamma(b_k)^{-1} .
\end{equation}

In its turn, the exponent of the double integral \eqref{Doub-intR} exhibits the following leading $T\to0$ behavior
 \begin{equation}\label{result-DI}
 \lim_{T\to0}\exp\left(\int\limits_{\mathbb{R}}
 \frac{z(\lambda)z(\mu)}{(\lambda-\mu_+)^2}\,d\lambda\,d\mu\right)
 \left(\frac{q\varepsilon'_0}{\pi T}\right)^{-\frac{u_1^2}{2\pi^2}}=e^{C_1\left[\frac{u_1(\lambda)}{2\pi i}\right]}
 G^2\Bigl(1,\frac{u_1}{2\pi i}\Bigr).
 \end{equation}
The functional $C_1[F]$ appearing above reads
 \begin{equation}\label{Def-C1}
 C_1[F(\lambda)]=\frac12\int\limits_{-q}^q\frac{F'(\lambda)F(\mu)-F(\lambda)F'(\mu)}{\lambda-\mu}
 \,d\lambda\,d\mu+2F(q)\int\limits_{-q}^q\frac{F(\lambda)-F(q)}{\lambda-q},
 \end{equation}
and $G(1,x)=G(1+x)G(1-x)$, where $G(x)$ is the Barnes function.

Now we substitute $u_1/2\pi i=\ell-\alpha_\ell{\cal Z}$. Combining
\eqref{det-Cauchy}, \eqref{gFp}, \eqref{prod-alpha}, and
\eqref{result-DI} and using
 \begin{equation}\label{C-C}
 C_1[\alpha_\ell Z(\lambda)-\ell]=C_1[\alpha_\ell Z(\lambda)]-4\ell \alpha_\ell
 \int\limits_{-q}^q\frac{Z(\mu)-{\cal Z}}{\mu-q}\,d\mu,
 \end{equation}
we obtain
 \begin{multline}\label{DC-1}
 B_d^{(0)}[u] = \lim_{T\to0} B_d[u]
 \left|\frac{q\varepsilon'_0}{i\pi T}\right|^{2\alpha_\ell ^2{ {\cal Z} }^2}
%
 =e^{C_1[\alpha_\ell Z(\lambda)]}\left(\frac{\sin\pi\alpha_\ell {\cal Z}}\pi\right)^{2n}G^2(1,\alpha_\ell {\cal Z}-\ell)\\
 \times R_{n_p^+,n_h^+}(\{p^+\},\{h^+\}|\alpha_\ell {\cal Z}-\ell) \; R_{n_p^-,n_h^-}(\{p^-\},\{h^-\}|\ell -\alpha_\ell {\cal Z}),
  \end{multline}
where
 \begin{equation}\label{R-def}
 R_{n,m}(\{p\},\{h\}|\nu)= \frac{\prod\limits_{j>k}^n(p_j-p_k)^2\prod\limits_{j>k}^m(h_j-h_k)^2}
 {\prod\limits_{j=1}^n\prod\limits_{k=1}^m(p_j+h_k-1)^2} \;
 \Gamma^2 \left( \begin{array}{c}
 \{p_k+\nu\}\  ,\ \{h_k-\nu\}\\
 \{p_k\}\  ,\ \{h_k\} \end{array}\right).
 \end{equation}

We have thus reproduced the discrete part of form factors
given in \cite{KitKMST10a} up to the
replacement $v_0/T\mapsto iL$.



\section{Final result\label{S-FR}}

We have calculated the low-temperature limits of the exponential
decay and constant coefficient in the long-distance asymptotic
expansion of the generating function $\langle e^{2i\pi \alpha {\cal
Q}_x }\rangle_T $ for given distribution of the roots
$\hat{s}_j^{\pm}$. In this section we sum up all the obtained
contributions for $\ell$ fixed.

Let us first summarize the results obtained in the previous
sections. The asymptotic expansion of the generating function
$\langle e^{2i\pi \alpha {\cal Q}_x }\rangle_T $ at low temperature
has the form
\begin{equation}\label{Q-Ql}
 \langle e^{2i\pi \alpha {\cal Q}_x }\rangle_T \simeq  \sum_{\ell \in \mathbb{Z} }^{}
 Q_\ell\; e^{2i\alpha_{\ell}k_{_{F}}x }\qquad x\to\infty,\quad T\to0, \quad xT\to\infty \;,
\end{equation}
where the coefficients $Q_\ell$ are
\begin{equation}\label{Ql}
 Q_\ell=B^{(0)}_s[\alpha_{\ell}Z] \left(\frac{\pi T}{q\varepsilon'_0}\right)^{2\alpha_\ell ^2{ {\cal
 Z} }^2}G^2(1,\alpha_\ell {\cal Z}-\ell)e^{C_1[\alpha_\ell
 Z(\lambda)]-\frac{2\pi Tx}{v_0}((\alpha_\ell{\cal Z})^2-\ell^2)}W_+W_-\;.
\end{equation}
Here the smooth part $B^{(0)}_s[\alpha_{\ell}Z]$ is given by
\eqref{AC2-W}, the functional $C_1$ by \eqref{Def-C1}. To describe
the factors $W_\pm$ we first define a function $W(\nu,r)$ depending
on complex $\nu$ and integer $r$ as
 \begin{multline}\label{H-def}
 W(\nu,r)=\sum_{n,n'=0\atop{n-n'=r}}^\infty\,\sum_{1\le
 p_1<\cdots<p_{n}<\infty}\,\sum_{1\le
 h_1<\cdots<h_{n'}<\infty}\prod_{j=1}^{n}e^{-\frac{2\pi Tx}{v_0}(p_j-1)}\prod_{k=1}^{n'}
 e^{-\frac{2\pi Tx}{v_0}h_k}\\
 \times \left(\frac{\sin\pi
 \nu}\pi\right)^{2n'}
 R_{n,n'}(\{p\},\{h\}|\nu),
 \end{multline}
where $R_{n,n'}(\{p\},\{h\}|\nu)$ is given by \eqref{R-def}. Then
 \begin{equation}\label{Hpm}
 W_\pm=\Bigl.W(\nu,r)\Bigr|_{\nu=\pm(\alpha_\ell{\cal
 Z}-\ell),\atop{r=\pm\ell}}\;,
 \end{equation}
It is easy to see that the factors $W_\pm$  correspond to the sums
over all the possible choices of integers $\{ p_{a}^{\pm}
\}_1^{n_p^{\pm}}$ and $\{ h_{a}^{\pm} \}_1^{n_h^{\pm}}$ which
parameterize the possible configurations of the roots
$\hat{s}_j^{\pm}$ at $\ell$ fixed. Such sums are computed in
\cite{KitKMST10b} as,
 \begin{equation}\label{magic-formula}
 W(\nu,r)=\frac{G^2(1+r+\nu)}{G^2(1+\nu)}\,\frac{e^{-\frac{\pi
 Tx}{v_0}r(r-1)}}{\Big(1-e^{-\frac{2\pi
 Tx}{v_0} }\Big)^{(\nu+r)^2}}.
 \end{equation}
Setting here $\nu=\pm(\alpha_\ell{\cal
 Z}-\ell)$, $r=\pm\ell$ and substituting \eqref{magic-formula} into
 \eqref{Ql} we obtain after simple algebra
\begin{equation}\label{Ql-new}
 Q_\ell= \widetilde{A}_{\ell }\left(\frac{\pi T/v_0}{\sinh\frac{\pi Tx}{v_0}}\right)^{2\alpha_\ell ^2{ {\cal
 Z} }^2},\quad\mbox{with}\quad \widetilde{A}_{\ell }=     B^{(0)}_s[\alpha_{\ell}Z] \frac{G^2(1,\alpha_\ell {\cal Z})}
 {(2q{\cal Z})^{2\alpha_\ell ^2{\cal
 Z}^2}}\; e^{C_1[\alpha_\ell
 Z(\lambda)]}\;,
\end{equation}
where we have used $\varepsilon'_0={\cal Z}v_0$. Thus we arrive at
the following asymptotic expansion
\begin{equation}\label{Asy-eaQ}
 \langle e^{2i\pi \alpha {\cal Q}_x }\rangle_T \simeq  \sum_{\ell \in \mathbb{Z} }^{} e^{2i\alpha_{\ell}k_{_{F}}x }
 \left(  \frac{ \pi T /v_0 }{\sinh\frac{\pi Tx}{v_0} }  \right)^{2 \alpha_{\ell}^2 \mathcal{Z}^2}   \widetilde{A}_{\ell }
\;.
\end{equation}

Note that the obtained result is obviously  a periodic function of
$\alpha$ as it was expected thanks to the fact that the coefficient $\widetilde{A}_{\ell }$ is also a function of ${\alpha_\ell}$ only, see appendix~\ref{SPoA-Sec}. It is also straightforward to see that
the combination $\widetilde A_{\ell}\ (\pi T/v_0)^{2 \ell^2
\mathcal{Z}^2}$ coincides with the amplitude of the critical form
factor of the operator $e^{2i\pi \alpha {\cal Q}_x }$
\cite{KitKMST10a} corresponding to the umklapp-type excited state of
the momentum $2\alpha_ \ell k_{{}_F}$, where $v_0/T$ plays the role
of the system size.

Finally, in order to obtain the long-distance asymptotic expansion
of the density-density correlation function it is enough to apply
the differential operator \eqref{8-cor-fun} to the equation
\eqref{Asy-eaQ}. Hereby one should distinguish two cases: $\ell=0$
and $\ell\ne 0$. In the last case one has due to \eqref{AC2-W}
 \begin{equation}\label{B-smooth}
 \Bigl.B^{(0)}_s[\alpha_\ell Z]\Bigr|_{\alpha=0}=0,\qquad
 \Bigl. \partial_\alpha B^{(0)}_s[\alpha_\ell Z]\Bigr|_{\alpha=0}=0.
 \end{equation}
Therefore the second $\alpha$-derivative should be applied to the
coefficient $B^{(0)}_s[\alpha_\ell Z]$.

On the contrary $B^{(0)}_s[\alpha_\ell Z]=1$ at $\alpha=0$ and
$\ell=0$, that is at $\alpha_\ell=0$, (see \cite{KozMS10}). Therefore in that case, the second
$\alpha$-derivative should be applied on the combination containing the dependence on $x$, namely on  
$\exp[2i\alpha_{\ell}k_{_{F}}x] \bigl(\sinh(\pi Tx/{v_0})
\bigr)^{-2\alpha_\ell ^2{\cal Z}^2}$, as otherwise the second
$x$-derivative vanishes. Thus, taking the second $\alpha$ and $x$
derivatives of \eqref{Asy-eaQ} as it is explained above, and neglecting higher order corrections over $T$ as well as subdominant exponential decays in $x$,  we arrive at
\eqref{CFT-SinhTerm} with
\begin{equation}
A_{\ell} = \frac{D^2 \ell ^2}{2}   \frac{
\partial^2 }{ \partial \alpha^2} \Bigl.  \widetilde{A}_{\ell }
\Bigr|_{\alpha=0} \;.
\end{equation}
It is readily checked that in the $x\to\infty$, $xT\to 0$  limit  
equation \eqref{CFT-SinhTerm} does reproduce the long-distance
asymptotic expansion of the density-density correlation function of
the one dimensional Bose gas at $T=0$, together with the correct
values of the amplitudes \cite{KitKMST09b}.



\section*{Acknowledgements}

We are very grateful to N. Kitanine and V. Terras for useful and
numerous discussions. J. M. M. and N. S.  are supported by CNRS.  We acknowledge the support from the GDRI-471 of CNRS
"French-Russian network in Theoretical and Mathematical  Physics"
and RFBR-CNRS-09-01-93106L-a. J. M. M. is also supported by the ANR grant DIADEMS 10 BLAN 012004, N. S.  by the Program
of RAS Mathematical Methods of the Nonlinear Dynamics,
RFBR-11-01-00440a, SS-8265.2010.1 and  K. K. K.  by the EU
Marie-Curie Excellence Grant MEXT-CT-2006-042695.  N. S. and K. K. K
would like to thank the Theoretical Physics group of the Laboratory
of Physics at ENS Lyon for hospitality, which makes this
collaboration possible.


\appendix

\section{Estimates of integrals with regular functions\label{SM-Int}}

In this appendix we estimate the class of integrals that appears in equations
\eqref{YY-orig} and \eqref{Inteq-u1}. We first focus on the integrals of the form:

 \begin{equation}\label{Int-f-or}
 J[\varepsilon]=T\int\limits_{\mathbb{R}}f(\lambda)\log\left(1+e^{-\frac{\varepsilon(\lambda)}T}\right)\,d\lambda.
 \end{equation}
For our purpose, it is enough to consider the case when $f(\lambda)$
is bounded on the real axis and differentiable in vicinities of $\pm
q$, although the result of  the analysis remains valid  at much less
restrictive assumptions. Due to the properties of
$\varepsilon(\lambda)$  \eqref{routh-est} it is clear that
 \begin{equation}\label{Int-f-or0}
 \lim_{T \to 0} J[\varepsilon]=-\int\limits_{-q}^qf(\lambda)\varepsilon(\lambda)\,d\lambda .
 \end{equation}
In order to find power-law corrections to the equation
\eqref{Int-f-or0}, one should estimate the contributions coming from
the vicinities of $\pm q$ more thoroughly. Let $\delta>0$ be such
that $\delta\to 0$ as $T\to 0$, while $\delta/T\to \infty$ as $T\to
0$. We can split the integral $J$ into five parts
$J=J_-+J_{-q}+J_0+J_q+J_+$. The integrals $J_\pm$ correspond to the
domains $\lambda>q+\delta$ and $\lambda<-q-\delta$. They behave as
$O(e^{-\varepsilon(\pm q \pm \delta)/T}) $ and hence produce
$O(T^{\infty})$ contributions.  The integral $J_0$ runs along the
domain $-q+\delta<\lambda<q-\delta$. By factoring out
$e^{-\frac{\varepsilon(\lambda)}{T}}$  from the logarithm, we get
that
 \begin{equation}\label{J0}
 J_0[\varepsilon]=-\int\limits_{-q+\delta}^{q-\delta}f(\lambda)\varepsilon(\lambda)\,d\lambda+O(T^\infty).
 \end{equation}

Finally the integrals $J_{\pm q}$ correspond to the domains $\pm
q-\delta<\lambda<\pm q+\delta$ and generate all power-law
corrections in T to \eqref{Int-f-or0}. We now derive the leading
power-law correction to \eqref{Int-f-or0} coming from the
$\delta$-vicinity of $q$. For doing this, we can replace the
functions entering the integral by the leading non-vanishing terms
of their Taylor expansions. Namely, we replace $f(\lambda)$ by
$f(q)$ and  $\varepsilon(\lambda)$ by
$(\lambda-q)\varepsilon'_0+T\varepsilon_1(q)$. Recall that
$\varepsilon_0(q)=0$ and we denote $\varepsilon'_0\equiv
\varepsilon'_0(q)$.  Thence,
 \begin{equation}\label{Jq1}
 J_q[\varepsilon]=Tf(q)\int\limits_{-\delta}^\delta
 \log\left(1+e^{-\frac{\lambda\varepsilon'_0}T-\varepsilon_1(q)}\right)\,d\lambda
 +h.o.c.,
 \end{equation}
where $h.o.c.$ means the higher order corrections in $T$. After
changing of variables $\lambda= \mu T/\varepsilon'_0$ we obtain
 \begin{multline}\label{Jq2}
 J_q[\varepsilon]=\frac{T^2f(q)}{\varepsilon'_0}\int\limits_{-\delta\varepsilon'_0/T}^{\delta\varepsilon'_0/T}
 \left[\log\left(1+e^{-\mu-\varepsilon_1(q)}\right)+\bigl(\mu+\varepsilon_1(q)\bigr)\Theta\bigl(-\mu-\varepsilon_1(q)\bigr)\right]\,d\mu\\
 -\frac{T^2f(q)}{\varepsilon'_0}\int\limits_{-\delta\varepsilon'_0/T}^{-\varepsilon_1(q)}
 \bigl(\mu+\varepsilon_1(q)\bigr)\,d\mu  +  h.o.c.,
 \end{multline}
where $\Theta(\lambda)$ is the Heaviside step-function. Using now
that $\delta/T\to \infty$ we arrive at
 \begin{multline}\label{Jq3}
 J_q[\varepsilon]=\frac{T^2f(q)}{2 \varepsilon'_0}
 \left(\varepsilon_1(q)-\frac{\delta\varepsilon'_0}T\right)^2+\frac{T^2f(q)}{\varepsilon'_0}\int\limits_{-\infty}^{\infty}
 \left[\log\left(1+e^{-\mu}\right)+\mu\Theta(-\mu)\right]\,d\mu
 \\
 =\frac{T^2f(q)}{2\varepsilon'_0}\left(\varepsilon_1(q)-\frac{\delta\varepsilon'_0}T\right)^2
 +\frac{\pi^2T^2f(q)}{6\varepsilon'_0} +  h.o.c. \; .
 \end{multline}
Similarly one has
 \begin{equation}\label{J-q1}
 J_{-q}[\varepsilon]=\frac{T^2f(-q)}{2\varepsilon'_0}\left(\varepsilon_1(-q)-\frac{\delta\varepsilon'_0}T\right)^2
 +\frac{\pi^2T^2f(-q)}{6\varepsilon'_0} +  h.o.c.  \; .
 \end{equation}
Combining \eqref{Jq3}, \eqref{J-q1} with \eqref{J0} we obtain after
simple algebra
 \begin{multline}\label{Intgr}
 J[\varepsilon]=-\int\limits_{-q}^qf(\lambda)\varepsilon(\lambda)\,d\lambda+
 \frac{T^2\pi^2}{6\varepsilon'_0}\bigl(
 f(q)+f(-q)\bigr)\\
 +\frac{T^2f(q)\varepsilon_1^2(q)}{2\varepsilon'_0}+\frac{T^2f(-q)\varepsilon_1^2(-q)}{2\varepsilon'_0}
 +  h.o.c. \; .
 \end{multline}

In a similar way, one can obtain the low-temperature expansion of integrals involving the function $u(\lambda)$
 \begin{equation}\label{Int-f-or-u}
 J[u]=T\int\limits_{\mathbb{R}}f(\lambda)\log\left(1+e^{-\frac{u(\lambda)}T}\right)\,d\lambda.
 \end{equation}
Since $u_0(\lambda)=\varepsilon_0(\lambda)$, exactly
the same considerations lead us to the estimate
 \begin{multline}\label{Intgr-u}
 J[u]=-\int\limits_{-q}^qf(\lambda)u(\lambda)\,d\lambda+
 \frac{T^2\pi^2}{6\varepsilon'_0}\bigl(
 f(q)+f(-q)\bigr)\\
 +\frac{T^2f(q)u_1^2(q)}{2\varepsilon'_0}+\frac{T^2f(-q)u_1^2(-q)}{2\varepsilon'_0}
 +O(T^3).
 \end{multline}

\section{Estimates of integrals with singular functions \label{sing-int}}

\subsection{The Cauchy transform in the vicinities of $\pm q$}

In this section we determine the leading $T\to 0$ behavior of $L_{\mathbb{R}}[z](\lambda)$. Its depends on where $\lambda$
is located. Recall that
 \begin{equation}\label{Init-Ca}
 L_{\mathbb{R}}[z](\lambda)=\frac{-1}{2\pi i}
 \int\limits_{\mathbb{R}}\log\left(\frac{1+e^{-\frac{u(\mu)}T}}{1+e^{-\frac{\varepsilon(\mu)}T}}\right)
 \frac{d\mu}{\mu-\lambda}.
 \end{equation}
If $\lambda$ is separated from $\pm q$, then obviously
 \begin{equation}\label{alpha-far}
 \lim_{T\to 0}  L_{\mathbb{R}}[z](\lambda)    =   \frac{1}{2\pi i}
 \int\limits_{-q}^q\frac{u_1(\mu)\,d\mu}{\mu-\lambda}=
 \frac{1}{2\pi i}L_{[-q,q]}[u_1](\lambda), \qquad
 \end{equation}
where $u_1(\lambda)$ is given by \eqref{u1}.

Let now $\lambda\to q$ as $T\to 0$. We denote $\lambda=\lambda_\pm$,
if $\lambda$  approaches $q$ from the upper (resp. lower)
half-plane. Let again $\delta>0$ be such that $\delta\to 0$ as $T\to
0$, while $\delta/T\to \infty$ in the $T\to 0$ limit. Consider the
contributions to the integral \eqref{Init-Ca} coming from different
intervals of integration. Obviously, when $T\to 0$ the integrals over domains
$\lambda>q+\delta$ and $\lambda<-q-\delta$ produce exponentially small
corrections. On the other hand $z(\lambda)$ can be approximated by
$\frac{u_1(\lambda)}{2\pi i}$ on the interval
$[-q-\delta,q-\delta]$:
 \begin{equation}\label{int-aux1}
 \frac{-1}{2\pi i}
 \int\limits_{-q-\delta}^{q-\delta}\log\left(\frac{1+e^{-\frac{u(\mu)}T}}{1+e^{-\frac{\varepsilon(\mu)}T}}\right)
 \frac{d\mu}{\mu-\lambda_\pm} \hookrightarrow   \frac{1}{2\pi
 i}\int\limits_{-q-\delta}^{q-\delta}\frac{u_1(\mu)\,d\mu}{\mu-\lambda_\pm},\qquad T\to
 0.
 \end{equation}
Extracting the divergent part we obtain
 \begin{equation}\label{div-part1}
 \frac{1}{2\pi
 i}\int\limits_{-q-\delta}^{q-\delta}\frac{u_1(\mu)\,d\mu}{\mu-\lambda_\pm}\to
 \frac{1}{2\pi
 i}\int\limits_{-q}^{q}\frac{u_1(\mu)-u_1}{\mu-\lambda_\pm}\,d\mu+
 \frac{u_1}{2\pi
 i}\log\left(\frac{\lambda_\pm-q+\delta}{\lambda_\pm+q}\right),\qquad
 \delta\to 0,
 \end{equation}
and we remind that $u_1=u_1(\pm q)$.

It remains to compute the integral over $[q-\delta,q+\delta]$.
Following the method of the previous section we linearize the
functions $u(\mu)$ and $\varepsilon(\mu)$ in the vicinity of
$\mu=q$. Then we have
 \begin{equation}\label{int-main1}
 I_q\equiv \frac{-1}{2\pi i}
 \int\limits_{q-\delta}^{q+\delta}\log\left(\frac{1+e^{-\frac{u(\mu)}T}}{1+e^{-\frac{\varepsilon(\mu)}T}}\right)
 \frac{d\mu}{\mu-\lambda_\pm} \hookrightarrow   \frac{-1}{2\pi i}
 \int\limits_{-\delta}^{\delta}\log\left(\frac{1+e^{-\frac{\mu\varepsilon'_0}T-u_1}}
 {1+e^{-\frac{\mu\varepsilon'_0}T}}\right)
 \frac{d\mu}{\mu-(\lambda_\pm-q)}.
 \end{equation}
Replacing $\mu\varepsilon'_0/T=\xi$ we arrive at
 \begin{equation}\label{int-main2}
 I_q=   \frac{-1}{2\pi
 i}\int\limits_{-\delta\varepsilon'_0/T}^{\delta\varepsilon'_0/T}\left[\log\left(\frac{e^{\xi}+e^{-u_1}}
 {e^{\xi}+1}\right)+u_1\Theta(-\xi-c)\right]
 \frac{d\xi}{\xi-t_\pm}+\frac{u_1}{2\pi i}\log\left(\frac{t_\pm+c}{t_\pm+\frac{\delta\varepsilon'_0}T}\right),
 \end{equation}
where $c$ is an arbitrary positive constant and we have set
$t_\pm=(\lambda_\pm-q)\varepsilon'_0/T$. We can send now
$\delta/T\to\infty$. Substituting into \eqref{int-main2}
 \begin{equation}\label{alpha-rep}
 \frac1{\xi-t_\pm}=\pm i\int\limits_0^\infty e^{\mp
 i\omega(\xi-t_\pm)}\,d\omega,
 \end{equation}
we arrive at
 \begin{multline}\label{int-main3}
 I_q=   \frac{\mp 1}{2\pi}\int\limits_{-\infty}^{\infty}\,d\xi\int\limits_{0}^{\infty}\,d\omega
 \left[\log\left(\frac{e^{\xi}+e^{-u_1}}
 {e^{\xi}+1}\right)+u_1\Theta(-\xi-c)\right]e^{\mp
 i\omega(\xi-t_\pm)}\\
 +\frac{u_1}{2\pi
 i}\log\left(\frac{t_\pm+c}{t_\pm+\frac{\delta\varepsilon'_0}T}\right).
 \end{multline}
The integral over $\xi$ can be calculated by means of an integration
by parts followed by a computation of the residues at
$e^{\xi}+e^{-u_1}=0$ and $e^{\xi}+1=0$:
 \begin{equation}\label{int-xi}
 \frac{1}{2\pi}\int\limits_{-\infty}^{\infty}\,d\xi
 \left[\log\left(\frac{e^{\xi}+e^{-u_1}}
 {e^{\xi}+1}\right)+u_1\Theta(-\xi-c)\right]e^{\mp
 i\omega\xi}=\frac{1-e^{\pm i\omega u_1}}{2\omega\sinh(\pi\omega)}
   \mp\frac{u_1 e^{\pm i\omega c}}{2\pi i\omega}.
   \end{equation}
Thus, we arrive at
 \begin{equation}\label{int-main4}
 I_q= \frac{u_1}{2\pi i}\log\left(\frac{t_\pm+c}{t_\pm+\frac{\delta\varepsilon'_0}T}\right)
 \pm \frac{1}{2\pi}\int\limits_{0}^{\infty}\frac{d\omega}{\omega}
 \left[\mp iu_1 e^{\pm i\omega
 c}-\frac{\pi}{\sinh(\pi\omega)}\left(1-e^{\pm i\omega
 u_1}\right) \right]e^{\pm
 i\omega t_\pm}.
 \end{equation}
Due to \eqref{Constr} the last integral is convergent. It  can be
computed in terms of the $\Gamma$-functions via
 \begin{equation}\label{2-int}
 \int\limits_0^\infty\frac{e^{-p\omega}\,d\omega}{\omega}\left[
 b-a-\frac{\pi}{\sinh(\pi\omega)}\left(e^{-a\omega}-e^{-b\omega}\right)\right]=
 (a-b)\log\left(\frac{p}{2\pi}\right)+2\pi\log
 \frac{\Gamma\left(\frac{p+b}{2\pi}+\frac12\right)}
 {\Gamma\left(\frac{p+a}{2\pi}+\frac12\right)}.
 \end{equation}
Thus, we obtain
 \begin{equation}\label{int-main5}
 I_q= \frac{u_1}{2\pi i}\log\left(\frac{\pm2\pi iT}{(\lambda_\pm-q+\delta)\varepsilon'_0}\right)
 \pm \log\frac{\Gamma\left(\frac12\pm\frac{(\lambda_\pm-q)\varepsilon'_0}
 {2\pi i T}\pm \frac{u_1}{2\pi i}\right)}
 {\Gamma\left(\frac12\pm\frac{(\lambda_\pm-q)\varepsilon'_0}
 {2\pi i T}\right)}.
 \end{equation}
Combining this result with \eqref{div-part1} we find the following
estimate
 \begin{multline}\label{alpha-+q+}
 L_{\mathbb{R}}[z](\lambda_\pm)=\frac{1}{2\pi i}
 \int\limits_{-q}^q\frac{u_1(\mu)-u_1}{\mu-\lambda_\pm}\,d\mu
 +\frac{u_1}{2\pi i}\log\left(\frac{\lambda_\pm-q}{\lambda_\pm+q}\right)\\
 -\frac{u_1}{2\pi i}\log\left(\frac{(\lambda_\pm-q)\varepsilon'_0}
 {\pm2\pi i T}\right)\pm\log\frac{\Gamma\left(\frac12\pm\frac{(\lambda_\pm-q)\varepsilon'_0}
 {2\pi i T}\pm \frac{u_1}{2\pi i}\right)}
 {\Gamma\left(\frac12\pm\frac{(\lambda_\pm-q)\varepsilon'_0}
 {2\pi i T}\right)},\qquad T\to 0,\quad \lambda\sim q.
 \end{multline}

Similarly, if $\lambda\to-q$ as $T\to0$ one has
 \begin{multline}\label{alpha--q+}
 L_{\mathbb{R}}[z](\lambda_\pm)=\frac{1}{2\pi i}
 \int\limits_{-q}^q\frac{u_1(\mu)-u_1}{\mu-\lambda_\pm}\,d\mu
 +\frac{u_1}{2\pi i}\log\left(\frac{\lambda_\pm-q}{\lambda_\pm+q}\right)\\
 +\frac{u_1}{2\pi i}\log\left(\frac{(\lambda_\pm+q)\varepsilon'_0}
 {\pm2\pi i T}\right)\pm\log\frac{\Gamma\left(\frac12\pm\frac{(\lambda_\pm+q)\varepsilon'_0}
 {2\pi i T}\mp \frac{u_1}{2\pi i}\right)}
 {\Gamma\left(\frac12\pm\frac{(\lambda_\pm+q)\varepsilon'_0}
 {2\pi i T}\right)},\qquad T\to 0,\quad \lambda\sim -q.
 \end{multline}

\subsection{The double integral}

Consider now the low temperature behavior of the integral A given in \eqref{Doub-intR-sing}. As usual we split the integration domain
into several pieces $A=A_- + A_{-q} + A_0 + A_{q}+A_+$. The integral $A_+$ (resp. $A_-$)
over the domain $\lambda>q+\delta$ (resp. $\lambda<-q-\delta$) are again exponentially small in respect to the $T\to 0$ limit.
When $\lambda \in [-q-\delta,q-\delta]$, we can use the expression \eqref{alpha-far}
for $L_{\mathbb{R}}[z](\lambda_+)$ and also replace $z(\lambda)$
by $\frac{u_1(\lambda)}{2\pi i}$. This gives
 \begin{equation}\label{Doub-cal1}
 A_0\equiv\frac1{(2\pi i)^2}\int\limits_{-q+\delta}^{q-\delta}\,d\lambda
  u_1(\lambda)\partial_\lambda\int\limits_{-q}^{q}\,d\mu
 \frac{u_1(\mu)}{\mu-\lambda_+}.
 \end{equation}
Integrating by parts we arrive at
 \begin{equation}\label{Doub-cal2}
 A_0=\frac{-1}{(2\pi i)^2}\int\limits_{-q+\delta}^{q-\delta}\,d\lambda
  \int\limits_{-q}^{q}\,d\mu
 \frac{u'_1(\lambda)u_1(\mu)}{\mu-\lambda_+}+
 \frac{u_1(q-\delta)}{(2\pi i)^2}\left(
 \int\limits_{-q}^{q}
 \frac{u_1(\mu)\,d\mu}{\mu_--q+\delta}-
 \int\limits_{-q}^{q}
 \frac{u_1(\mu)\,d\mu}{\mu_-+q-\delta}\right).
 \end{equation}
Here we have used that $u_1(\lambda)=u_1(-\lambda)$. This last
property also allows one to symmetrize the integrand, so that upon sending $\delta  \to 0$, we get
 \begin{equation}\label{Doub-Doub}
 \frac{-1}{(2\pi i)^2}\int\limits_{-q+\delta}^{q-\delta}\,d\lambda
  \int\limits_{-q}^{q}\,d\mu
 \frac{u'_1(\lambda)u_1(\mu)}{\mu-\lambda_+}=
 \frac{1}{2(2\pi i)^2}\int\limits_{-q}^{q}
 \frac{u'_1(\lambda)u_1(\mu)-u_1(\lambda)u'_1(\mu)}{\lambda-\mu}\,d\lambda\,d\mu,
 \qquad \delta\to 0.
 \end{equation}
Extracting then the divergent part from the single integrals in
\eqref{Doub-cal2} we find
 \begin{equation}\label{Doub-cal3}
 A_0\to C_1\left[\frac{u_1(\lambda)}{2\pi
 i}\right]+\frac{2u_1^2}{(2\pi i)^2}\log\left(\frac{\delta}{2q}\right),\qquad \delta\to 0,
 \end{equation}
where the functional $C_1[F]$ is defined in \eqref{Def-C1}.

Consider now the contribution $A_q$ coming from the interval $
q-\delta<\lambda< q+\delta$:
 \begin{equation}\label{Aq-1}
 A_q=\int\limits_{q-\delta}^{q+\delta}
 \bigl(\partial_\lambda L_{\mathbb{R}}[z](\lambda_+)\bigr)\cdot
 z(\lambda)\,d\lambda.
 \end{equation}
Substituting here \eqref{alpha-+q+} we arrive at
$A_q=A_q^{(1)}+A_q^{(2)}+A_q^{(3)}$, where
 \begin{equation}\label{Aq1}
 A_q^{(1)}=\frac1{2\pi i}\int\limits_{q-\delta}^{q+\delta}z(\lambda)
 \partial_\lambda\left[\int\limits_{-q}^q\frac{u_1(\mu)-u_1}{\mu-\lambda_\pm}\,d\mu
 -\frac{u_1}{2\pi i}\log(\lambda_\pm+q)\right]\,d\lambda,
 \end{equation}
 \begin{equation}\label{Aq2}
 A_q^{(2)}=\frac{-u_1}{ (2\pi i)^2 }
 \int\limits_{-q-\delta}^{q+\delta}\log\left(\frac{1+e^{-\frac{u(\lambda)}T}}{1+e^{-\frac{\varepsilon(\lambda)}T}}\right)
 \frac{d\lambda}{\lambda_+-q} ,
 \end{equation}
and
 \begin{multline}\label{Aq3}
 A_q^{(3)}=\frac{-1}{2\pi i}
 \int\limits_{-q-\delta}^{q+\delta}
 \partial_\lambda\left\{\log\frac{\Gamma\left(\frac12+\frac{(\lambda_+-q)\varepsilon'_0}
 {2\pi i T}+ \frac{u_1}{2\pi i}\right)}
 {\Gamma\left(\frac12+\frac{(\lambda_+-q)\varepsilon'_0}
 {2\pi i T}\right)}-\frac{u_1}{2\pi i}\log\left(\frac{(\lambda_+-q)\varepsilon'_0}
 {2\pi i T}\right)\right\}\\
 \times \log\left(\frac{1+e^{-\frac{u(\lambda)}T}}{1+e^{-\frac{\varepsilon(\lambda)}T}}\right)\,d\lambda.
 \end{multline}

It is easy to see that $A_q^{(1)}\to 0$ as $\delta\to 0$, because
the integrand  is a bounded function as $T\to 0$. The integral
$A_q^{(2)}$ can be estimated similarly to \eqref{int-main1}:
 \begin{equation}\label{Aq2-est}
 A_q^{(2)}=\frac{u_1^2}{(2\pi i)^2}\log\left(\frac{-2\pi i T}{\delta\varepsilon'_0}\right)
 -\frac{u_1}{2\pi i}\log\frac{\Gamma\left(\frac12-\frac{u_1}{2\pi i}\right)}
 {\Gamma\left(\frac12\right)},\qquad T\to 0,\quad \delta\to 0.
 \end{equation}
As for the remaining integral $A_q^{(3)}$, its leading behavior is
obtained by a linearization of the functions $u(\lambda)$ and
$\varepsilon(\lambda)$ in the vicinity of $\lambda=q$. After the
change of variables $\xi = (\lambda-q)\varepsilon'_0/ T$ followed by
an integration by parts, we find  in the $\delta\to 0$, $T \to 0$
limit
 \begin{equation}\label{Aq3-2}
 A_q^{(3)}
 =\frac{1}{2\pi i} \int\limits_{-\infty}^{\infty}
 \left\{\log\frac{\Gamma\left(\frac12+\frac{\xi+u_1}
 {2\pi i }\right)}
 {\Gamma\left(\frac12+\frac{\xi}
 {2\pi i }\right)}-\frac{u_1}{2\pi i}\log\left(\frac{\xi+i0}
 {2\pi i }\right)\right\}
 \left(\frac1{1+e^{-\xi -u_1}}-\frac1{1+e^{-\xi}}\right)\,d\xi .
 \end{equation}
We  close the integration contour in the upper half-plane and
compute the integral \eqref{Aq3-2} by residues. These are located at
$\xi=-u_1+\pi i(2k+1)$ and $\xi=\pi i(2k+1)$, $k=0,1\dots$. Hence,
 \begin{multline}\label{Aq3-3}
 A_q^{(3)}=\sum_{k=1}^\infty\left[\log\frac{\Gamma^2(k)}{\Gamma\left(k+
 \frac{u_1} {2\pi i }\right)\Gamma\left(k- \frac{u_1} {2\pi i
 }\right)}-\frac{u_1} {2\pi i }\log\left(\frac{k-\frac12-\frac{u_1} {2\pi i
 }}{k-\frac12}\right)\right]\\
 =\log G\Bigl(1,\frac{u_1} {2\pi i }\Bigr)+\frac{u_1} {2\pi i
 } \log\frac{\Gamma\left(\frac12-\frac{u_1}{2\pi i}\right)}
 {\Gamma\left(\frac12\right)}\\
 +\lim_{N\to\infty}\left[\log\frac{G^2(N+1)}{G\left(N+1+
 \frac{u_1} {2\pi i }\right)G\left(N+1- \frac{u_1} {2\pi i
 }\right)}-\frac{u_1} {2\pi i }\log\frac{\Gamma\left(N+\frac12-\frac{u_1} {2\pi i
 }\right)}{\Gamma\left(N+\frac12\right)}\right],
 \end{multline}
where $G(x)$ is the Barnes function and $G(1,x)=G(1+x)G(1-x)$. Using
the asymptotic behavior of the $\Gamma$ and Barnes functions for  $z
\to \infty$ with  $ z \not\in \mathbb{R}_-$
\begin{equation}
\begin{array}{l}
\log G(z+1+a) -\log G(z+1)   = a \log \sqrt{2\pi} + \frac{a}{2}
(2z+a) \log z - a z + \text{o}(1),\\
\log\Gamma(z+1+a) -\log \Gamma(z+1)  = a \log z + \text{o}(1),
\end{array}
\nonumber
\end{equation}
we find that the limit in the last line of \eqref{Aq3-3}
vanishes. Hence,
 \begin{equation}\label{Aq3-4}
 A_q^{(3)}=\log G\Bigl(1,\frac{u_1} {2\pi i }\Bigr)+\frac{u_1} {2\pi i
 } \log\frac{\Gamma\left(\frac12-\frac{u_1}{2\pi i}\right)}
 {\Gamma\left(\frac12\right)}.
 \end{equation}
Combining this result with \eqref{Aq2-est} we find
 \begin{equation}\label{Aq2-Aq3}
 A_q\to \frac{u_1^2}{(2\pi i)^2}\log\left(\frac{-2\pi i T}{\delta\varepsilon'_0}\right)
 +\log G\Bigl(1,\frac{u_1} {2\pi i }\Bigr),\qquad T\to 0,\quad \delta\to 0.
 \end{equation}

Similar calculation in the vicinity of the point $-q$ leads us to
the following below contribution coming from the interval $
q-\delta<\lambda< q+\delta$:
 \begin{equation}\label{A-q}
 A_{-q}\equiv \int\limits_{-q-\delta}^{-q+\delta}
 \bigl(\partial_\lambda L_{\mathbb{R}}[z](\lambda_+)\bigr)\cdot
 z(\lambda)\,d\lambda\to
 \frac{u_1^2}{(2\pi i)^2}\log\left(\frac{2\pi i T}{\delta\varepsilon'_0}\right)
 +\log G\Bigl(1,\frac{u_1} {2\pi i }\Bigr),\quad T\to 0,\quad \delta\to 0.
 \end{equation}
Thus, taking into account \eqref{Aq2-Aq3}, \eqref{A-q} and
\eqref{Doub-cal3} we finally obtain
 \begin{equation}\label{Doub-cal4}
 A\to C_1\left[\frac{u_1(\lambda)}{2\pi
 i}\right]-2\left(\frac{u_1}{2\pi i}\right)^2
 \log\left(\frac{q\varepsilon'_0}{\pi T}\right)+
 2\log G\Bigl(1,\frac{u_1}{2\pi i}\Bigr),\qquad T\to 0.
 \end{equation}

\section{Smooth part of the amplitude\label{SPoA-Sec}}

In this section we give the exact expression for the smooth part of
the amplitude $B^{(0)}_s[\alpha_{\ell}Z]= \lim_{T\to 0} B_s[u]$.
Provided the condition \eqref{sum-n} holds, we have
 \begin{multline}\label{AC2-W}
 B^{(0)}_s[ \alpha_{\ell}Z ]=(e^{2\pi i\alpha}-1)^2e^{ -C_0}\frac{\det\left(I+\frac1{2\pi i}\hat U^{(1)}[\alpha_{\ell}Z]\right)
 \det\left(I+\frac1{2\pi i} \hat
 U^{(2)}[ \alpha_{\ell}Z ]\right)}{\left(\det\left[I-{\textstyle\frac1{2\pi}}K\right]\right)^2}\numa{35}
 \times \left(e^{ -\alpha_\ell L[Z](\theta_1+i c)}  -e^{2\pi
 i\alpha-\alpha_\ell L[Z](\theta_1-i c)}\right)^{-1} \left(e^{ \alpha_\ell L[Z](\theta_2-i c)}
 -e^{2\pi
 i\alpha+\alpha_\ell L[Z](\theta_2+ic)}\right)^{-1}.
 \end{multline}
Here $L[Z](\omega)$ stands the Cauchy transform of the dressed
charge $Z(\lambda)$ on the interval $[-q,q]$, and $C_0$ is given by
\eqref{7-C0-lim}. The integral operator
$I-{\textstyle\frac1{2\pi}}K$ acts on the interval $[-q,q]$ and its
kernel was defined by \eqref{K-th}.  The operators
$I+\frac1{2\pi i}\hat U^{(1)}[ \alpha_{\ell}Z ]$ and $I+\frac1{2\pi i}\hat U^{(2)}[ \alpha_{\ell}Z ]$ act on a
anticlockwise oriented closed contour surrounding  $[-q,q]$. Their
kernels are
 \begin{equation}\label{AC2-U1}
 \hat U^{(1)}(w,w',[ \alpha_{\ell}Z ])=-e^{ -\alpha_\ell L[Z](w)}\cdot
 \frac{K_\alpha(w-w')-K_\alpha(\theta_1-w')}{
 e^{ -\alpha_\ell L[Z](w+i c)}  -e^{2\pi i\alpha-\alpha_\ell L[Z](w-i c)}},
 \end{equation}
and
 \begin{equation}\label{AC2-U2}
 \hat U^{(2)}(w,w',[ \alpha_{\ell}Z ])=e^{ \alpha_\ell L[Z](w')}\cdot
 \frac{K_\alpha(w-w')-K_\alpha(w-\theta_2)}{
 e^{ \alpha_\ell L[Z](w'-i c)}  -e^{2\pi i\alpha+\alpha_\ell L[Z](w'+i
 c)}},
 \end{equation}
where $K_\alpha(\lambda)$ is given by \eqref{Kalpha}. Finally
parameters $\theta_1$ and $\theta_2$ are arbitrary complex numbers
lying inside of the contour where the operators $\hat
U^{(1,2)}(w,w',[ \alpha_{\ell}Z ])$ act. If we set $\theta_1=-q$ and $\theta_2=q$,
then we reproduce the smooth part of form factors of the $\mathbf{P}_\ell$
class obtained \cite{KitKMST10a}.

\end{document}